\documentclass[twocolumn,tighten,times]{aastex61}
\shorttitle{Star Formation in the V380 Ori NE Region}
\shortauthors{Choi et al.}

\submitjournal{To appear in the Astrophysical Journal Supplement Series}

\begin{document}

\fontsize{10}{10.6}\selectfont

\title{Precessing Jet and Large Dust Grains
       in the V380 Ori NE Star-forming Region}
       
\author{Minho Choi}
\affiliation{Korea Astronomy and Space Science Institute,
             776 Daedeokdaero, Yuseong, Daejeon 34055, Korea}

\author{Miju Kang}
\affiliation{Korea Astronomy and Space Science Institute,
             776 Daedeokdaero, Yuseong, Daejeon 34055, Korea}

\author{Jeong-Eun Lee}
\affiliation{School of Space Research, Kyung Hee University,
             1732 Deogyeong-daero, Giheung-gu, Yongin-si,
             Gyeonggi-do 17104, Korea}

\author{Ken'ichi Tatematsu}
\affiliation{National Astronomical Observatory of Japan,
             National Institutes of Natural Sciences, 2-21-1 Osawa,
             Mitaka, Tokyo 181-8588, Japan}
\affiliation{Department of Astronomical Science,
             SOKENDAI (The Graduate University for Advanced Studies),
             2-21-1 Osawa, Mitaka, Tokyo 181-8588, Japan}

\author{Sung-Ju Kang}
\affiliation{Korea Astronomy and Space Science Institute,
             776 Daedeokdaero, Yuseong, Daejeon 34055, Korea}

\author{Jack Sayers}
\affiliation{Department of Physics, Math, and Astronomy,
             California Institute of Technology, Pasadena, CA 91125, USA}

\author{Neal J. Evans II}
\affiliation{Korea Astronomy and Space Science Institute,
             776 Daedeokdaero, Yuseong, Daejeon 34055, Korea}
\affiliation{Department of Astronomy, University of Texas at Austin,
             2515 Speedway, Stop C1400, Austin, TX 78712-1205, USA}

\author{Jungyeon Cho}
\affiliation{Department of Astronomy and Space Science,
             Chungnam National University, 99 Daehak-ro, Yuseong-gu,
             Daejeon, 34134, Korea}

\author{Jungmi Kwon}
\affiliation{Institute of Space and Astronautical Science,
             Japan Aerospace Exploration Agency, 3-1-1 Yoshinodai,
             Chuo-ku, Sagamihara, Kanagawa 252-5210, Japan}

\author{Geumsook Park}
\affiliation{Korea Astronomy and Space Science Institute,
             776 Daedeokdaero, Yuseong, Daejeon 34055, Korea}
\affiliation{Department of Astronomy and Space Science,
             Chungnam National University, 99 Daehak-ro, Yuseong-gu,
             Daejeon, 34134, Korea}
\affiliation{Department of Physics and Astronomy,
             Seoul National University, Seoul 151-747, Korea}

\author{Satoshi Ohashi}
\affiliation{The Institute of Physical and Chemical Research (RIKEN),
             2-1 Hirosawa, Wako-shi, Saitama 351-0198, Japan}

\author{Hyunju Yoo}
\affiliation{School of Space Research, Kyung Hee University,
             1732 Deogyeong-daero, Giheung-gu, Yongin-si,
             Gyeonggi-do 17104, Korea}
\affiliation{Department of Astronomy and Space Science,
             Chungnam National University, 99 Daehak-ro, Yuseong-gu,
             Daejeon, 34134, Korea}

\author{Youngung Lee}
\affiliation{Korea Astronomy and Space Science Institute,
             776 Daedeokdaero, Yuseong, Daejeon 34055, Korea}

\begin{abstract}
The V380 Ori NE bipolar outflow was imaged
in the SiO and CO $J$ = 1 $\rightarrow$ 0 lines,
and dense cores in L1641 were observed in the 2.0--0.89 mm continuum.
The highly collimated SiO jet shows point-symmetric oscillation patterns
in both position and velocity,
which suggests that the jet axis is precessing
and the driving source may belong to a non-coplanar binary system.
By considering the position and velocity variabilities together,
accurate jet parameters were derived.
The protostellar system is viewed nearly edge-on,
and the jet has a flow speed of $\sim$35 km s$^{-1}$
and a precession period of $\sim$1600 years.
The CO outflow length gives a dynamical timescale of $\sim$6300 years,
and the protostar must be extremely young.
The inferred binary separation of 6--70 au implies
that this protobinary system may have been formed
through the disk instability process.
The continuum spectra of L1641 dense cores indicate
that the emission comes from dust,
and the fits with modified blackbody functions
give emissivity power indices of $\beta$ = 0.3--2.2.
The emissivity index shows
a positive correlation with the molecular line width,
but no strong correlation with bolometric luminosity or temperature.
V380 Ori NE has a particularly low value of $\beta$ = 0.3,
which tentatively suggests the presence of millimeter-sized dust grains.
Because the dust growth takes millions of years,
much longer than the protostellar age,
this core may have produced large grains in the starless core stage.
HH 34 MMS and HH 147 MMS also have low emissivity indices.
\end{abstract}


\section{INTRODUCTION}

\enlargethispage{-1\baselineskip}

L1641 is an active site of star formation,
located in the southern part of the Orion A giant molecular cloud.
The L1641 region contains
thousands of low- and intermediate-mass young stellar objects (YSOs)
but no high-mass stars (Strom et al. 1989, 1993; Hsu et al. 2012, 2013).
Submillimeter continuum surveys revealed about 60 dense cores
in the L1641 cloud, many of them containing protostars
(Johnstone \& Bally 2006; Nutter \& Ward-Thompson 2007).
Furlan et al. (2016) identified and characterized
more than a hundred protostars in L1641.
Some of them have unusually red spectral energy distributions (SEDs)
and may be extremely young protostars (Stutz et al. 2013).

The ongoing star formation activities in L1641
are manifested by numerous YSO outflows
(Morgan et al. 1991; Stanke et al. 2002; Davis et al. 2009).
Levreault (1988) found a CO bipolar outflow
located to the northeast of V380 Ori and named it V380 Ori NE,
but could not identify the driving source.
Later studies showed that this outflow is driven
by a deeply embedded protostar (Davis et al. 2000),
one of the extreme Class 0 protostars reported by Stutz et al. (2013).
This protostar, HOPS 169, has
a bolometric luminosity of $L_{\rm bol}$ = 3.9 $L_\odot$
and a bolometric temperature of $T_{\rm bol}$ = 33 K (Furlan et al. 2016).
Maps of the infrared H$_2$ line
and the CO $J$ = 4 $\rightarrow$ 3 line revealed
that the V380 Ori NE outflow is a highly collimated jet
flowing along a point-symmetric curve
(Davis et al. 2000; Stanke et al. 2002).
Davis et al. (2000) suggested
that the outflow variability may be caused
by either jet deflection by dense ambient cloud
or precession of the driving source.
In addition to the very cold SED and highly collimated variable jet,
the V380 Ori NE outflow is unusually bright in the CH$_3$OH lines
(Kang et al. 2013).
These characteristics warrant detailed studies of the V380 Ori NE system,
which can help to understand the structure and activities of protostars
in the earliest stage of star formation.

Gibb et al. (2004) surveyed protostellar outflows
in the SiO $J$ = 5 $\rightarrow$ 4 line
and found that V380 Ori NE shows the SiO emission only at high velocities.
The lack of emission from the ambient cloud makes it possible
to study the detailed properties of the outflow without ambiguity.
These properties are in turn useful
for probing the physics of the outflow-launching region,
namely the inner part of the accretion disk.
Studies of SiO jets in other star forming regions
have indeed been fruitful in understanding the star formation activities,
such as the mass-ejection history, shock propagation,
jet collimation, jet bending, and jet rotation
(Dutrey et al. 1997; Choi 2005; Cabrit et al. 2007;
Codella et al. 2007; Choi et al. 2011a; Lee et al. 2015).

In this paper, we present the results of
our observations of the V380 Ori NE region with
the Very Large Array (VLA)
of the National Radio Astronomy Observatory (NRAO),
the Taeduk Radio Astronomy Observatory (TRAO),
and the Caltech Submillimeter Observatory (CSO).
We describe the observations in Section 2.
In Section 3, we report the results of the SiO imaging with VLA,
molecular line spectroscopy and mapping with TRAO,
and the millimeter--submillimeter continuum imaging with CSO.
In Section 4, we discuss
the star-forming activities of V380 Ori NE and other nearby objects.
A summary is given in Section 5.

\section{OBSERVATIONS}

\subsection{VLA}

The V380 Ori NE region was observed using the VLA
in the SiO $v=0$ $J$ = 1 $\rightarrow$ 0 line (43.423858 GHz).
Twenty-two antennas were used in the D-array configuration
on 2006 January 5.
The spectral window was set to have 32 channels
with a channel width of 0.39 MHz,
giving a velocity resolution of 2.7 km s$^{-1}$.

The nearby quasar 0541--056 (QSO B0539--057) was observed
to determine the phase and to obtain the bandpass response.
The flux was calibrated by observing the quasar 0319+415 (3C 84)
and setting its flux density to 7.92 Jy.
A comparison of the amplitude gave a flux density of 1.26 Jy for 0541--056.
To avoid any degradation of the sensitivity owing to pointing errors,
the pointing was checked
by observing the calibrators at the X band ($\lambda$ = 3.6 cm).
This pointing was performed
about once an hour, and just before observing the flux calibrator.

A two-field mosaic technique was used.
The phase tracking centers were ($\alpha_{2000}$, $\delta_{2000}$)
= (05$^{\rm h}$36$^{\rm m}$36\fs1, --06$^\circ$38$'$37\farcs4)
and (05$^{\rm h}$36$^{\rm m}$36\fs1, --06$^\circ$39$'$08\farcs6)
for the first and the second fields, respectively.
The angular distance between the field centers (31\farcs2)
corresponds to the half width of the primary beam.
Each field was imaged using a CLEAN algorithm.
With a natural weighting,
the visibility data of the first and the second fields
produced synthesized beams
of 2\farcs3 $\times$ 1\farcs5 and 2\farcs2 $\times$ 1\farcs5, respectively,
in full width at half-maximum (FWHM).
At the end of the CLEAN process, the maps were restored
using a circular Gaussian beam of FWHM = 1\farcs9 for both fields.
The maps were corrected for the primary beam response
and combined to produce a mosaic map.
The intensity of the mosaic map was then tapered
toward the edge of the field of view
to make the noise level roughly uniform across the map.
The imaging was also performed with a robust weighting,
using a restoring beam of FWHM = 1\farcs6.
The tapered mosaic maps have
rms noise levels of $\sim$1.4 mJy beam$^{-1}$ for the natural weighting
and $\sim$1.5 mJy beam$^{-1}$ for the robust weighting.
There is no continuum emission at a detectable level.

Another protostellar outflow, HH 340/343,
was observed in the same observing track,
to study the precessing jet driven by IRAS 03256+3055 (Hodapp et al. 2005).
The phase-tracking center was
(03$^{\rm h}$28$^{\rm m}$45\fs3, 31$^\circ$05$'$42\farcs0),
and the phase calibrator was 0336+323 (PKS 0333+321).
The bootstrapped flux density of 0336+323 was 0.79 Jy.
The image noise level of the HH 340/343 region
is similar to that of the V380 Ori NE region.
The SiO line was not detected.
We omit further description of the HH 340/343 region in this paper. 

\subsection{TRAO}

\begin{deluxetable}{ccc}[t]
\tabletypesize{\small}
\tablecaption{\small TRAO Telescope Parameters}%
\tablewidth{0pt}
\tablehead{
\colhead{Frequency\tablenotemark{a}} & \colhead{Beam FWHM}
& \colhead{Efficiency} \\
\colhead{(GHz)} & \colhead{(arcsec)} & \colhead{$\eta_{\rm mb}$} }%
\startdata
\phn86.24 & 60 & 0.46 \\
\phn98.00 & 53 & 0.52 \\
   115.27 & 45 & 0.51 \\
\enddata
\tablenotetext{a}{Frequency where the beam size and main-beam efficiency
                  were measured.}%
\end{deluxetable}

The V380 Ori NE region was observed
using the Second Quabbin Optical Imaging Array
(SEQUOIA-TRAO; Erickson et al. 1999)
receiver system on the TRAO 14 m telescope.
The data were obtained in 2016 January--February,
during an initial test period after the receiver array was relocated
from the Five College Radio Astronomy Observatory.
The backend consisted of an FFT spectrometer.
The spectral window has 4096 channels in a 62.5 MHz bandwidth.
The telescope parameters are listed in Table 1.
Telescope pointing was checked by observing Orion IRc2 (Baudry et al. 1995)
in the SiO $v$ = 1 $J$ = 2 $\rightarrow$ 1 maser line.
The pointing observations were performed about once in an hour.
The typical pointing error was $\sim$8$''$.
The antenna temperature was calibrated by the standard chopper-wheel method,
which automatically corrected for the effects of atmospheric attenuation.
We will present the TRAO data
in the main-beam temperature ($T_{\rm mb}$) scale.

The central source was observed
in the H$^{13}$CO$^+$ $J$ = 1 $\rightarrow$ 0 line (86.754294 GHz)
and the C$^{34}$S $J$ = 2 $\rightarrow$ 1 line (96.412982 GHz)
using the reference pixel of the receiver array.
The spectral channel width corresponds to a velocity width
of 0.053 and 0.047 km s$^{-1}$, respectively.
Typical system temperatures were 200 and 180 K, respectively.
The target position was ($\alpha_{2000}$, $\delta_{2000}$)
= (05$^{\rm h}$36$^{\rm m}$36\fs1, --06$^\circ$38$'$53$''$),
which corresponds to the submillimeter continuum source (Davis et al. 2000).
The data were taken by position switching
with the reference position of
($\Delta\alpha$, $\Delta\delta$) = (--40$'$, --40$'$)
relative to the target position.
For each spectrum, a first-order baseline was removed.
The spectral baseline was determined from the velocity intervals
of (3, 5) and (9, 11) km s$^{-1}$.

The 10$'$ $\times$ 10$'$ region around the target position
was mapped in the CO $J$ = 1 $\rightarrow$ 0 line (115.271204 GHz).
The typical system temperature was $\sim$580 K.
The data were taken by on-the-fly mapping
(see Jackson et al. 2006 for details of the mapping mode)
with the reference position of
($\Delta\alpha$, $\Delta\delta$) = (--90$'$, --90$'$).
Scans along the right ascension and declination were alternated.
The data were regridded onto a 25$''$ grid covering the mapping region.
For each spectrum, a first-order baseline was removed.
The spectral baseline was determined from the velocity intervals
of (--21, --7) and (21, 35) km s$^{-1}$.
One of the aims of the CO mapping was to check the data reliability.
The resulting CO map and spectra are consistent
with those presented by Morgan et al. (1991).

\subsection{CSO}

The northern part of the L1641 cloud was observed
using the CSO 10.4 m telescope in 2014 November--December.
The data were obtained using the Multiwavelength Submillimeter
Inductance Camera (MUSIC; Golwala et al. 2012; Sayers et al. 2014, 2016)
during a commissioning run of the instrument.
Detailed observational parameters are listed in Table 2.

\begin{deluxetable}{lccccc}[t]
\tabletypesize{\small}
\tablecaption{Summary of the MUSIC Parameters}%
\tablewidth{0pt}
\tablehead{
\colhead{Band} & \colhead{Wavelength} & \colhead{Beam}
& \colhead{$\epsilon$\tablenotemark{a}}
& \colhead{$\eta_{\rm hpf}$\tablenotemark{b}}
& \colhead{$\sigma$\tablenotemark{c}} \\
& \colhead{(mm)} & \colhead{FWHM} &&& \colhead{(Jy beam$^{-1}$)}}%
\startdata
0 & 2.0\phn & 48$''$ & 6.0\% & 0.89 & 0.10    \\
1 & 1.4\phn & 36$''$ & 4.6\% & 0.92 & 0.11    \\
2 & 1.1\phn & 32$''$ & 5.5\% & 0.93 & 0.2\phn \\
3 & 0.89    & 29$''$ & 9.4\% & 0.94 & 1.0\phn \\
\enddata
\tablecomments{See Sayers et al. (2016) for more details.}%
\tablenotetext{a}{Uncertainty in the absolute flux calibration.}%
\tablenotetext{b}{Correction factor for the high-pass filter,
                  i.e., the ratio of the peak source amplitude
                  in the filtered map
                  compared with the nominal peak source amplitude.}%
\tablenotetext{c}{Noise rms level in each map.}%
\end{deluxetable}

The data were taken by scanning the sky in a raster pattern.
Scans along the right ascension and declination were alternated.
Four wavelength bands were observed simultaneously.
The target mapping region was a 60$'$ $\times$ 60$'$ square
centered at ($\alpha_{2000}$, $\delta_{2000}$)
= (05$^{\rm h}$36$^{\rm m}$25\fs4, --06\arcdeg42$'$59$''$).
Uranus was observed once each night for flux and pointing calibrations.
The pointing accuracy was $\sim$5$''$.

The calibration and imaging were performed by the MUSIC commissioning team
(Sayers et al. 2016).
The data were processed with a high-pass filter
to reduce low-frequency noise,
which removes signals contained in Fourier modes larger than $\sim$16$'$.
For the sources in our maps, which are generally smaller than 40$''$,
this effect of the high-pass filter is minimal
and can be approximated by the correction factor $\eta_{\rm hpf}$
listed in Table 2.
(See Section 3.3 for the effect of the filtering on extended sources.)
The map of each wavelength band was divided by $\eta_{\rm hpf}$
to correct for the effect of the high-pass filter.

The correction factors were determined
by creating mock images of a point-like source,
matching the point spread function in each band.
These images were reverse-mapped into time-ordered data
using a scan pattern matching the real observations.
The mock data were processed identically to the real time-ordered data,
and maps were produced.
The peak heights of the source in these maps
were compared to those in the original mock images
to obtain the $\eta_{\rm hpf}$ values.

\section{RESULTS}

\subsection{VLA SiO Imaging}

\begin{figure}[!t]
\epsscale{1.0}
\plotone{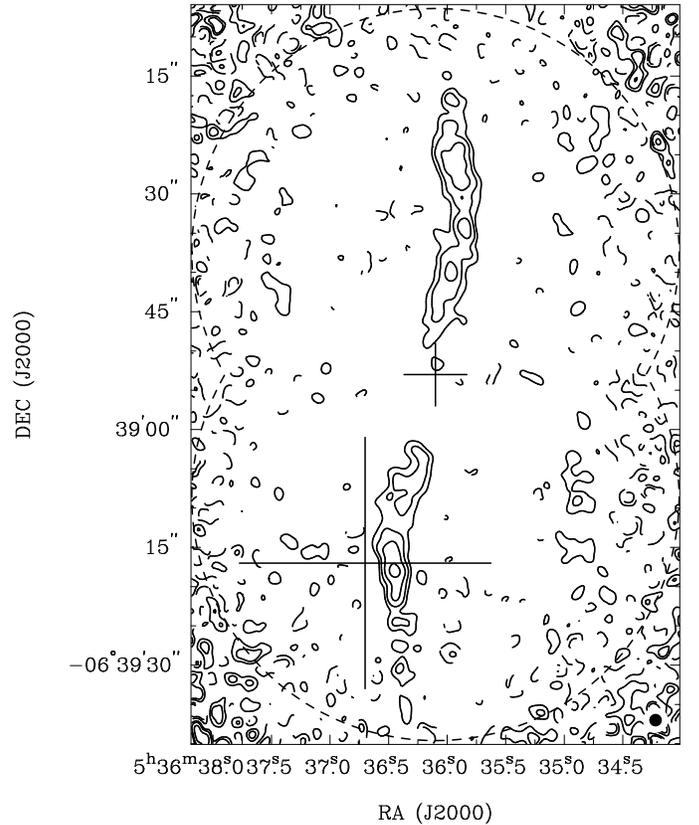}
\caption{
Map of the SiO $v$ = 0 $J$ = 1 $\rightarrow$ 0 line toward V380 Ori NE,
made with a robust weighting
and averaged over the velocity range
of $V_{\rm LSR}$ = (--9.0, 23.3) km s$^{-1}$.
Contour levels are 1, 2, 4, and 8 $\times$ 1.2 mJy beam$^{-1}$,
and the rms noise is 0.4 mJy beam$^{-1}$ near the map center.
Dashed contours are for negative levels.
The intensity taper was not applied,
and the noise level is higher at the edge of the map than at the center.
Shown in the bottom right-hand corner is the restoring beam: FWHM = 1\farcs6.
The background heat-scale image
shows the 4.5 $\mu$m band map from {\it Spitzer Space Telescope}.
Small plus sign:
the submillimeter continuum peak (Davis et al. 2000).
Large plus sign:
the CH$_3$OH emission peak KLC 8 (Kang et al. 2013).
Dashed arcs:
primary beam at each mosaic field.
[No background image in the arXiv version]}
\end{figure}

\begin{figure*}[!t]
\epsscale{1.0}
\plotone{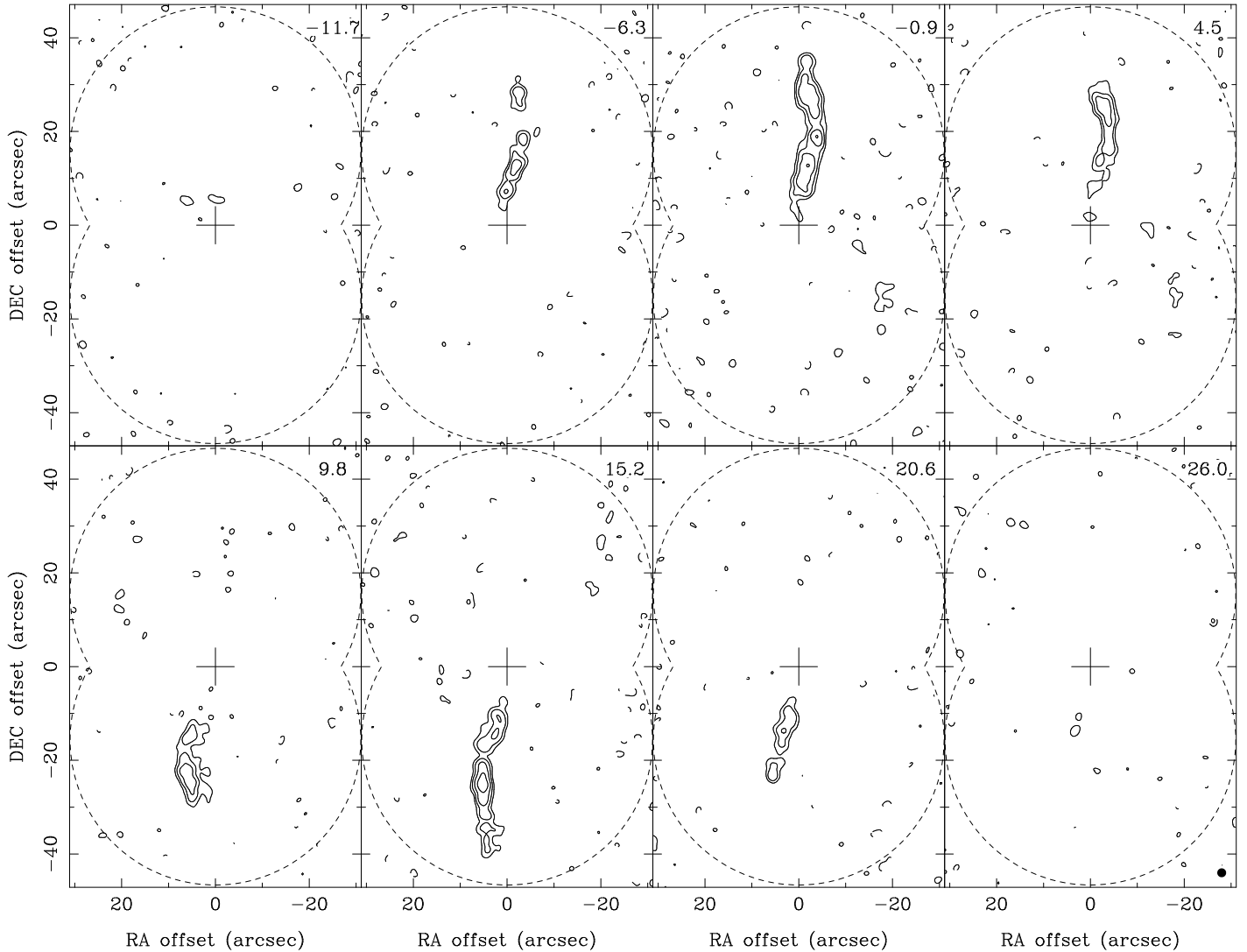}
\caption{
Velocity channel maps of the SiO line toward V380 Ori NE,
made with a natural weighting.
Central velocity of each map is written in the top right-hand corner,
and the velocity width for each map is 5.4 km s$^{-1}$.
Contour levels are 1, 2, 4, and 8 $\times$ 3 mJy beam$^{-1}$,
and the rms noise is 1.0 mJy beam$^{-1}$.
The intensity is tapered toward the edge of the field of view.
Shown in the bottom right-hand corner is the restoring beam: FWHM = 1\farcs9.
Plus sign:
the submillimeter continuum peak (Davis et al. 2000).}
\end{figure*}

The SiO line emission map (Figure 1) shows
the structure of the V380 Ori NE outflow system in the field of view.
The highly collimated bipolar jet flows along a gentle curve
that clearly shows a point-symmetric morphology.
The overall shape of the SiO jet is
consistent with that of the H$_2$ jet (Davis et al. 2000)
and the 4.5 $\mu$m emission features (Figure 1).
The velocity channel maps (Figure 2) show
that the northern jet is entirely blueshifted
and the southern jet is entirely redshifted.
This clear separation between the blueshifted and redshifted gas suggests
that the outflow collimation angle is smaller
than the inclination angle of the outflow axis
with respect to the plane of the sky.

\begin{figure}[!t]
\epsscale{1.0}
\plotone{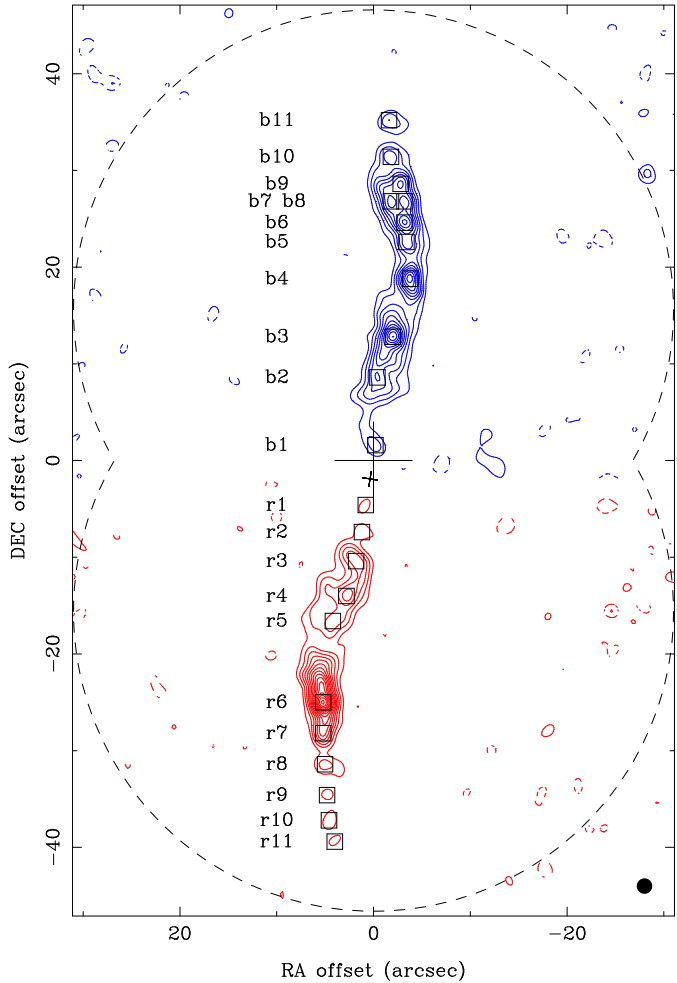}
\caption{
Maps of the SiO line toward V380 Ori NE,
made with a robust weighting.
In the northern half of the map,
the blue contours show the SiO line intensity
averaged over the velocity range of (--9.0, 7.2) km s$^{-1}$.
In the southern half of the map,
the red contours show the intensity averaged over (7.2, 23.3) km s$^{-1}$.
The lowest contour level is 1.8 mJy beam$^{-1}$,
and the contour interval is 1.2 mJy beam$^{-1}$.
The intensity is tapered toward the edge of the field of view.
Shown in the bottom right-hand corner is the restoring beam: FWHM = 1\farcs6.
Plus sign:
the submillimeter continuum peak (Davis et al. 2000).
Small cross:
expected position of the protostar (Figure 4).
Squares:
peak positions of the SiO emission.}
\end{figure}

At a smaller scale,
the SiO jet consists of a series of emission peaks (Figure 3).
The northern jet shows a relatively even intensity distribution.
That is, the strongest three peaks (b3, b4, and b6)
have comparable intensities.
By contrast, in the southern jet,
the strongest peak (r6) dominates the intensity distribution.
The intensity distribution of the H$_2$ jet
shows a similar asymmetry (Davis et al. 2000).
The northern H$_2$ jet shows a relatively continuous flow,
while the southern H$_2$ jet can be seen only around the brightest knots.
The SiO map shows the outflow structure near the driving source
much better than the H$_2$ image,
probably because the H$_2$ line is subject to extinction.
The spatial distribution of the SiO outflow peaks
shows an undulation pattern.
The position angle of the outflow peaks with respect to the driving source
seems to oscillate periodically.
This pattern is as expected for a precessing outflow.
Each of the northern/southern jets in the SiO image
covers more than a full cycle of the oscillation.

\begin{figure}[!t]
\epsscale{1.0}
\plotone{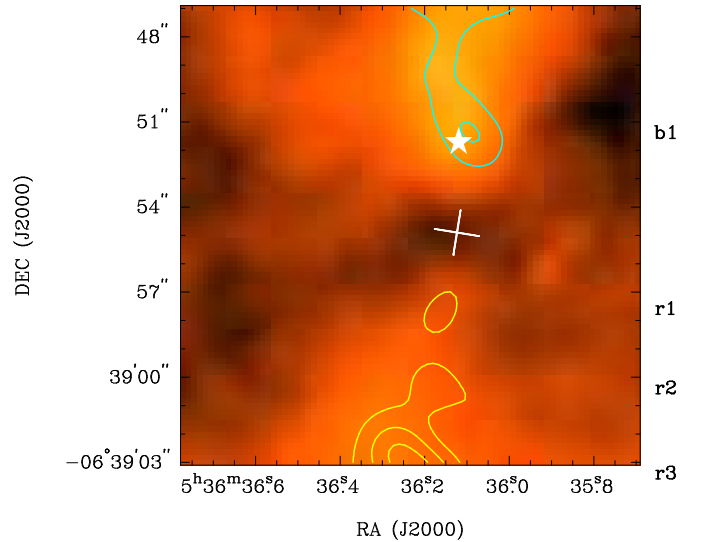}
\caption{
Maps of the SiO line, as shown in Figure 3,
in the central 16$''$ $\times$ 16$''$ region.
Contour levels are 1, 2, and 3 $\times$ 1.8 mJy beam$^{-1}$.
The SiO outflow peaks are labeled on the right edge of the figure.
The background heat-scale image
shows the 3.6 $\mu$m band map from {\it Spitzer Space Telescope},
with the extended nebulosity subtracted out.
Cross:
position of the protostar
inferred from the outflow axis and infrared dark lane.
Star symbol:
HOPS 169 (Furlan et al. 2016).}
\end{figure}

To describe the directional variability of the SiO jet
and analyze the oscillation of the position angle,
it is necessary to accurately determine the position of the driving source.
Among the emission sources known so far,
the best indicator for the position of the central protostar
is the submillimeter continuum source (Davis et al. 2000),
but the angular resolution ($\sim$14$''$) is much worse
than that of the SiO image.
The near-infrared position of HOPS 169  given by Furlan et al. (2016)
corresponds to an emission knot of the northern outflow (b1),
not the protostar itself.
Examinations of the near-infrared image (Figure 4) show
that there is a local intensity minimum
near the midpoint between the outflow peaks b1 and r1.
This feature is elongated in the east-west direction.
It seems to be a dark lane
caused by the edge-on disk or flattened envelope structure
around the protostar.
The intersection between the infrared dark lane
and the SiO outflow axis (b1--r1 line)
is a good candidate for the protostellar position.
The coordinate of this position is 
(05$^{\rm h}$36$^{\rm m}$36\fs12, --06$^\circ$38$'$54\farcs9).

\begin{figure}[!t]
\epsscale{1.0}
\plotone{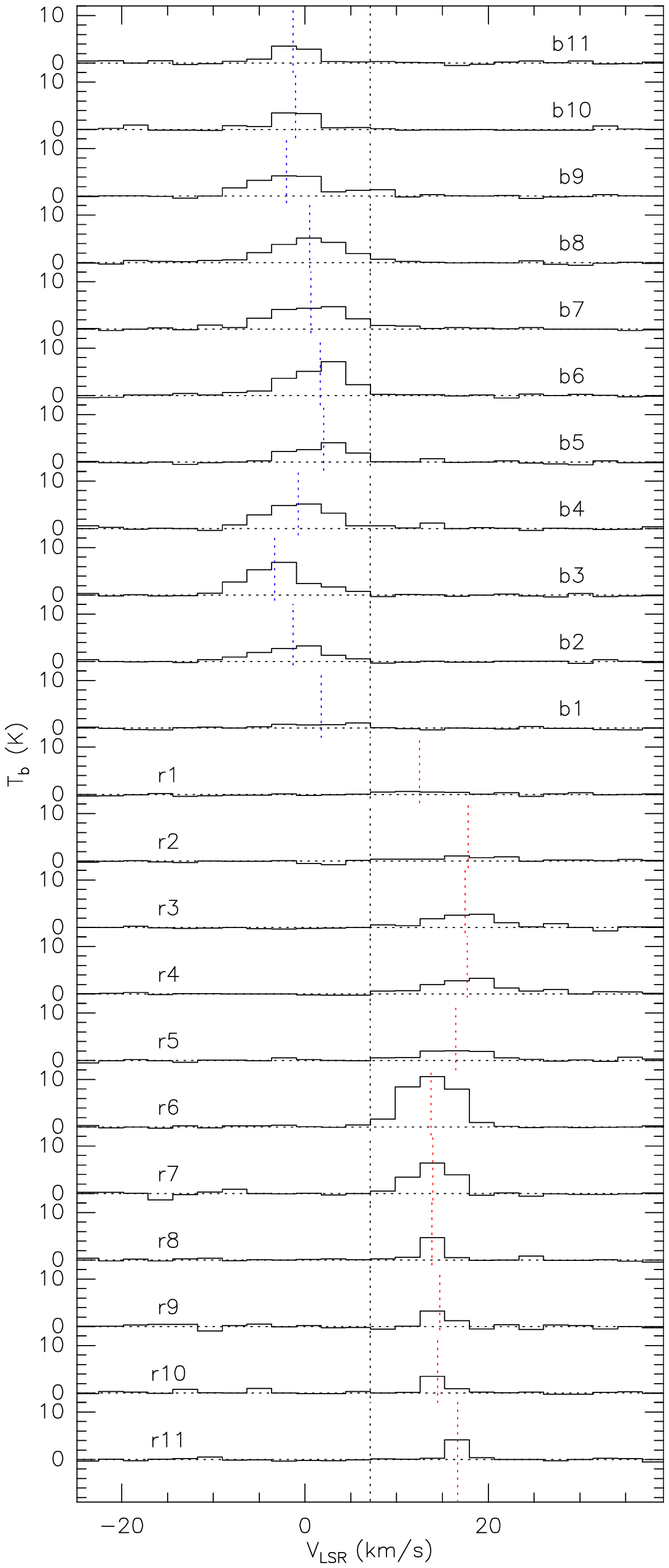}
\caption{
Spectra of the SiO line at the outflow peaks
(see Figure 3 for the positions of the outflow peaks).
The spectra were taken from the map made with a natural weighting.
The intensity taper was not applied.
Vertical black dotted line:
systemic velocity of V380 Ori NE,
$V_{\rm LSR}$ = 7.1 km s$^{-1}$ (Section 3.2).
Blue/red dotted lines:
centroid velocity.}
\end{figure}

\begin{figure*}[!t]
\epsscale{0.7}
\plotone{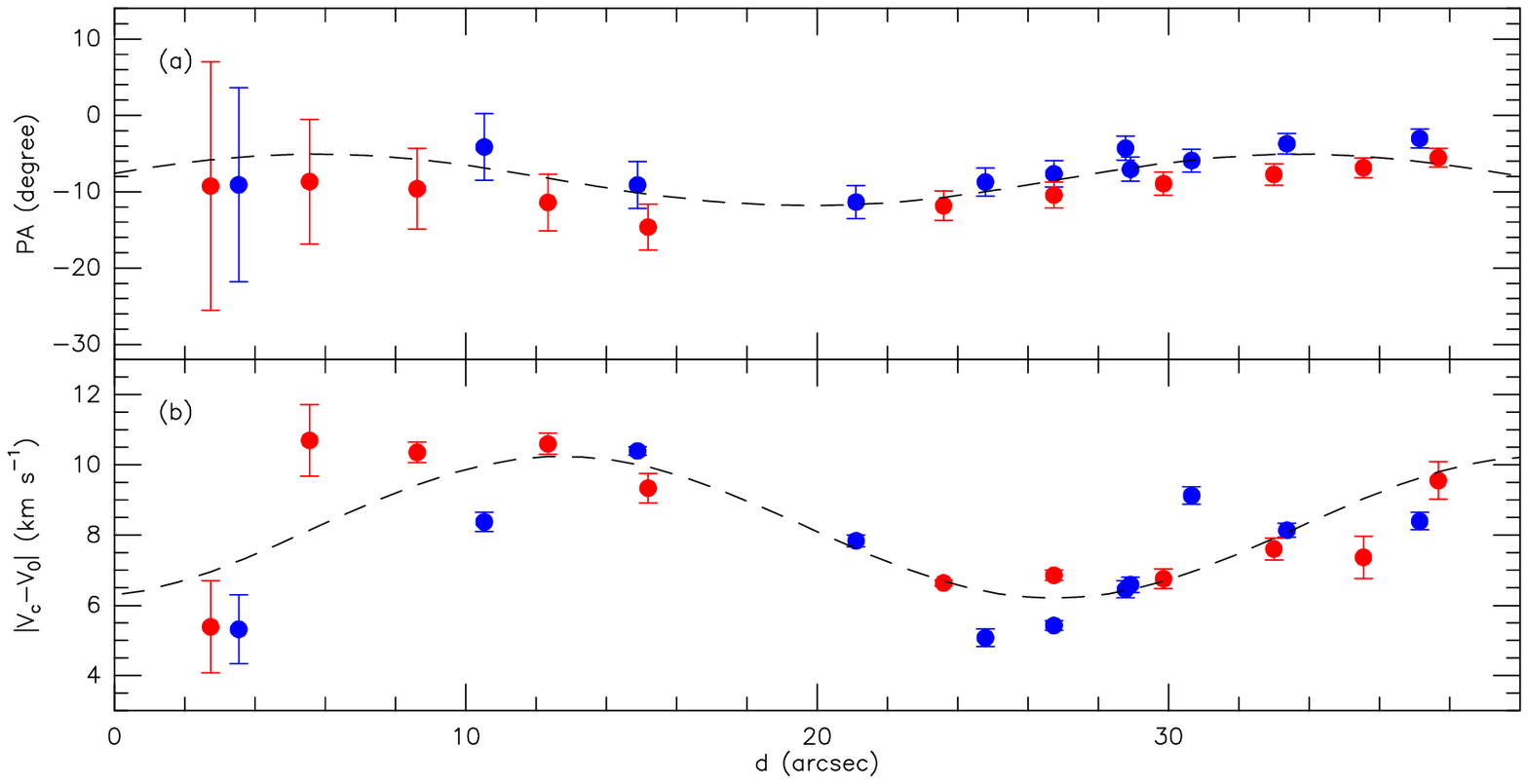}
\caption{
Parameters of the SiO emission peaks as functions of angular distance
from the expected position of the protostar.
Blue and red dots are for the northern and southern jets, respectively.
(a)
Position angle with respect to the protostellar position.
For the southern jet, the plot shows the position angle minus 180$^\circ$.
Vertical bars show the angle subtended by the beam (1\farcs6).
(b)
Centroid velocity with respect to the systemic velocity.
Vertical bars show the uncertainty of the centroid velocity
from the Gaussian fit to each spectrum.
Dashed curves:
best-fit sinusoidal functions.}
\end{figure*}

Figure 5 shows the SiO spectra at the outflow peaks.
Centroid velocities were derived
by fitting each spectrum with a single Gaussian profile.
Along each of the northern/southern outflows,
the centroid velocity shows an undulation pattern.
The spatial scale (wavelength) of the velocity undulation
is similar to that of the position angle (Figure 6).
The centroid velocity distribution also shows
an obvious point symmetry against the driving source.
The undulation patterns and point symmetries
of the position angle and centroid velocity suggest
that the directional variability and the velocity variability
are closely related phenomena, most likely having a single root cause.
The precession of jet axis may be a natural explanation.
Here we use the term ``precessing jet''
to describe the oscillating and point-symmetric variability of the jet,
which does not necessarily mean
that it is caused by the precession of the disk (Section 4.3).

\subsection{TRAO Molecular Line Spectroscopy and Mapping}

\begin{figure}[!t]
\epsscale{1.0}
\plotone{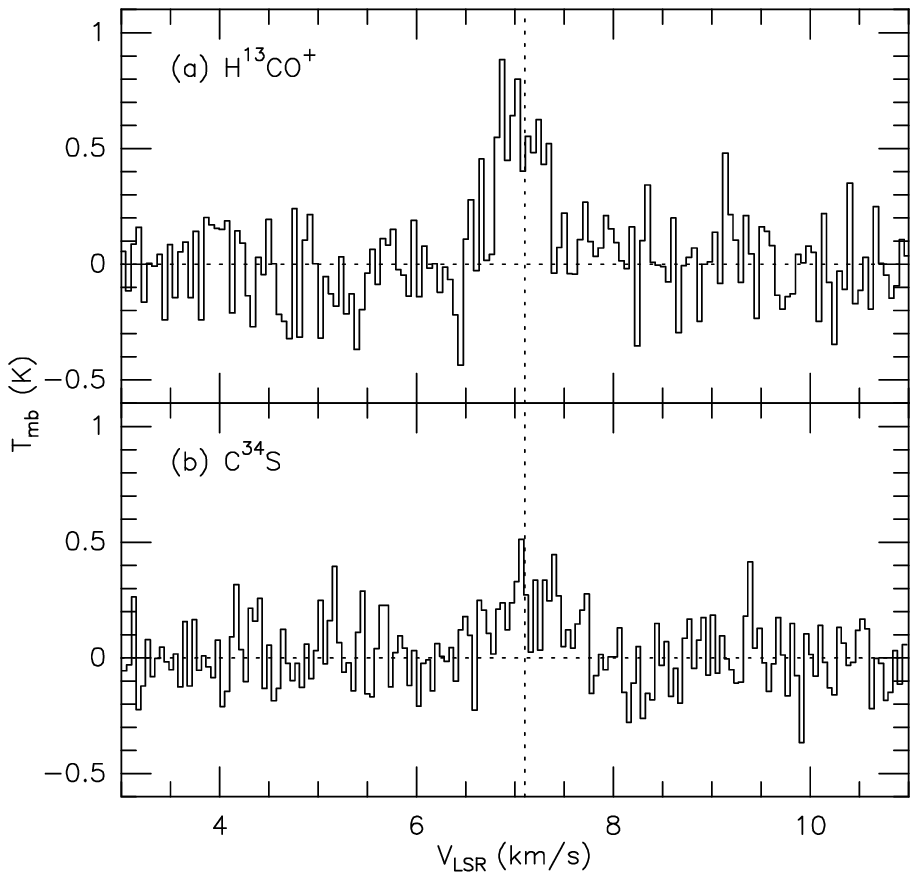}
\caption{
Spectra of (a) the H$^{13}$CO$^+$ $J$ = 1 $\rightarrow$ 0 line
and (b) the C$^{34}$S $J$ = 2 $\rightarrow$ 1 line
toward the V380 Ori NE dense core.
Vertical dotted line:
$V_{\rm LSR}$ = 7.1 km s$^{-1}$.}
\end{figure}

The H$^{13}$CO$^+$ and C$^{34}$S lines were observed
to determine the systemic velocity of the dense cloud core
containing the V380 Ori NE protostar.
These lines are expected to be optically thin.
Figure 7 shows the spectra.
Both lines were detected clearly.
The H$^{13}$CO$^+$ line especially shows a well-defined single peak
with a good signal-to-noise ratio.
The best-fit Gaussian profiles give centroid velocities
of 7.05 $\pm$ 0.03 km s$^{-1}$ for H$^{13}$CO$^+$
and 7.16 $\pm$ 0.06 km s$^{-1}$ for C$^{34}$S.
The line widths are FWHM = 0.55 and 0.78 km s$^{-1}$, respectively.
These velocities are consistent
with that of the C$^{18}$O $J$ = 3 $\rightarrow$ 2 line (Wilson et al. 1999).
Therefore, the systemic velocity of the V380 Ori NE dense core
is determined to be $V_0$ = 7.1 km s$^{-1}$.

\begin{figure}[!t]
\epsscale{1.0}
\plotone{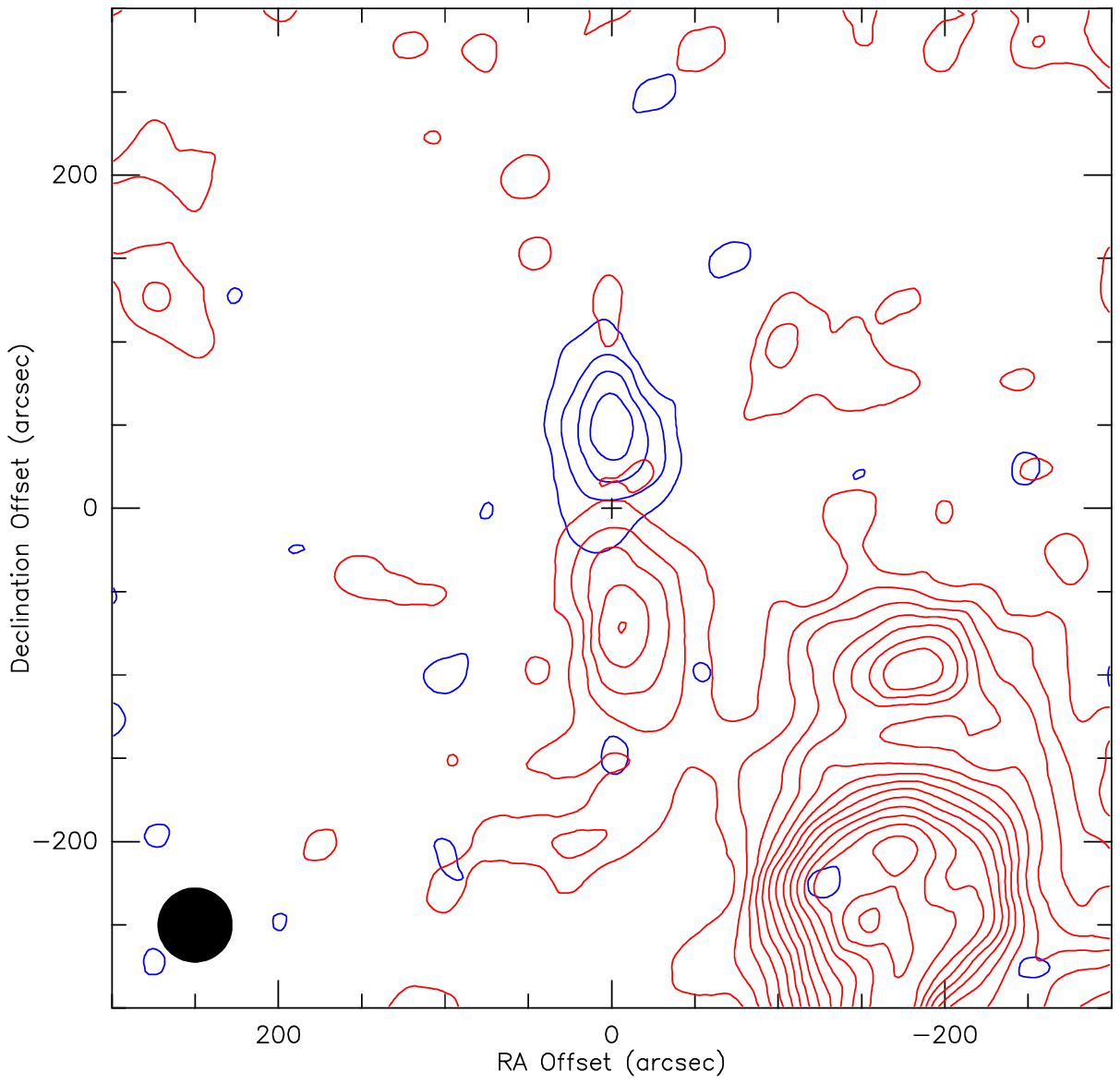}
\caption{
Maps of the CO $J$ = 1 $\rightarrow$ 0 line toward V380 Ori NE.
The blue and red contours show the CO line intensity
integrated over the velocity ranges
of (--3.7, 4.1) and (10.0, 20.4) km s$^{-1}$, respectively.
The lowest contour levels are 2 K km s$^{-1}$ and 3 K km s$^{-1}$
for blue and red contours, respectively.
The contour interval is 2 K km s$^{-1}$.
Shown in the bottom left-hand corner is the beam: FWHM = 45$''$.
Plus sign:
the submillimeter continuum peak (Davis et al. 2000),
which is the map center.}
\end{figure}

The CO line was observed to measure the total extent
of the V380 Ori NE bipolar outflow.
Figure 8 shows the maps of line wings.
These maps have a denser sampling and cover a larger area
than those presented by Morgan et al. (1991).
The length of each outflow lobe is $\sim$140$''$.
Considering the beam size,
the deconvolved length of the outflow may be $\sim$120$''$.
There is no sign of outflow beyond this length.
The CO emission component in the southwestern corner of Figure 8
corresponds to the cloud associated with NGC 1999.
As in the case of the SiO outflow,
the CO outflow also shows a clear separation
between the blueshifted and redshifted gas,
which suggests that the CO outflow must be well-collimated.

\subsection{CSO Continuum Imaging}

Initial inspections of the resulting MUSIC images showed
that several emission sources are detectable.
Comparison of the source positions in the MUSIC maps
with those of submillimeter counterparts (Nutter \& Ward-Thompson 2007)
revealed that the position difference is $\sim$4$''$ on average,
which is consistent with the pointing accuracy.

\begin{figure}[!t]
\epsscale{1.0}
\plotone{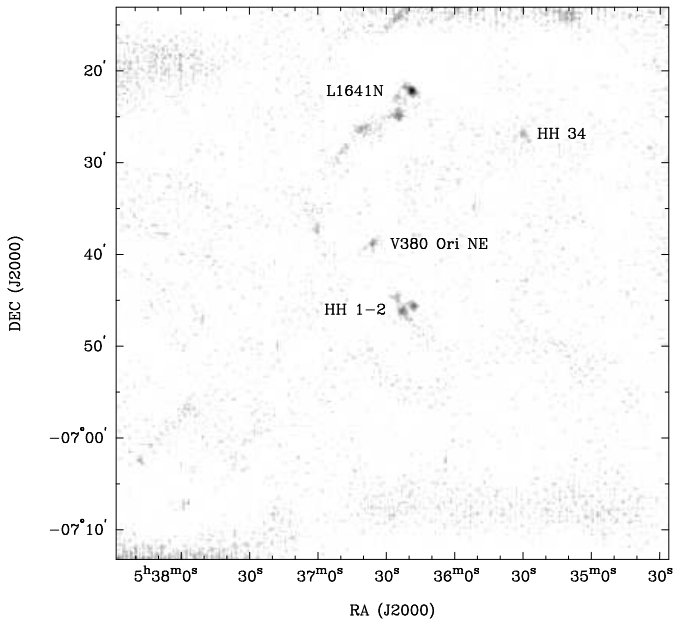}
\caption{
Combined image of the $\lambda$ = 1.4 and 1.1 mm continuum,
showing the 60$'$ $\times$ 60$'$ region of the MUSIC map.
The maps of the two bands were averaged
with a weighting to maximize the signal-to-noise ratio.
[Low-quality image in the arXiv version]}
\end{figure}

The MUSIC maps match the sensitivity of similar maps in the literature
(Johnstone \& Bally 2006; Nutter \& Ward-Thompson 2007; Davis et al. 2009),
but provide data in multiple bands,
enabling the study of spectral properties.
We searched for sources detectable in at least two bands.
For a source with a typical spectral slope of dust continuum emission,
the signal-to-noise ratio is higher in the $\lambda$ = 1.4 and 1.1 mm bands
than in the other bands.
Figure 9 shows the combined map of the 1.4 and 1.1 mm bands.
The noise levels are not entirely uniform over the mapping region,
and there are some patches of relatively high noise.
Eight sources were identified,
and their properties are listed in Table 3.
There are probable sources near the northern edge of the map,
but they are ignored because the noise level is high around them.
Some of the detected objects, such as L1641N MMS,
contain several protostars and may be considered as clumps,
while others, such as V380 Ori NE,
contain a single protostar (or a protobinary)
and may be considered as dense cloud cores.
In this paper, we refer to the detected objects as ``cores'' for simplicity.

\begin{figure*}[!t]
\epsscale{0.9}
\plotone{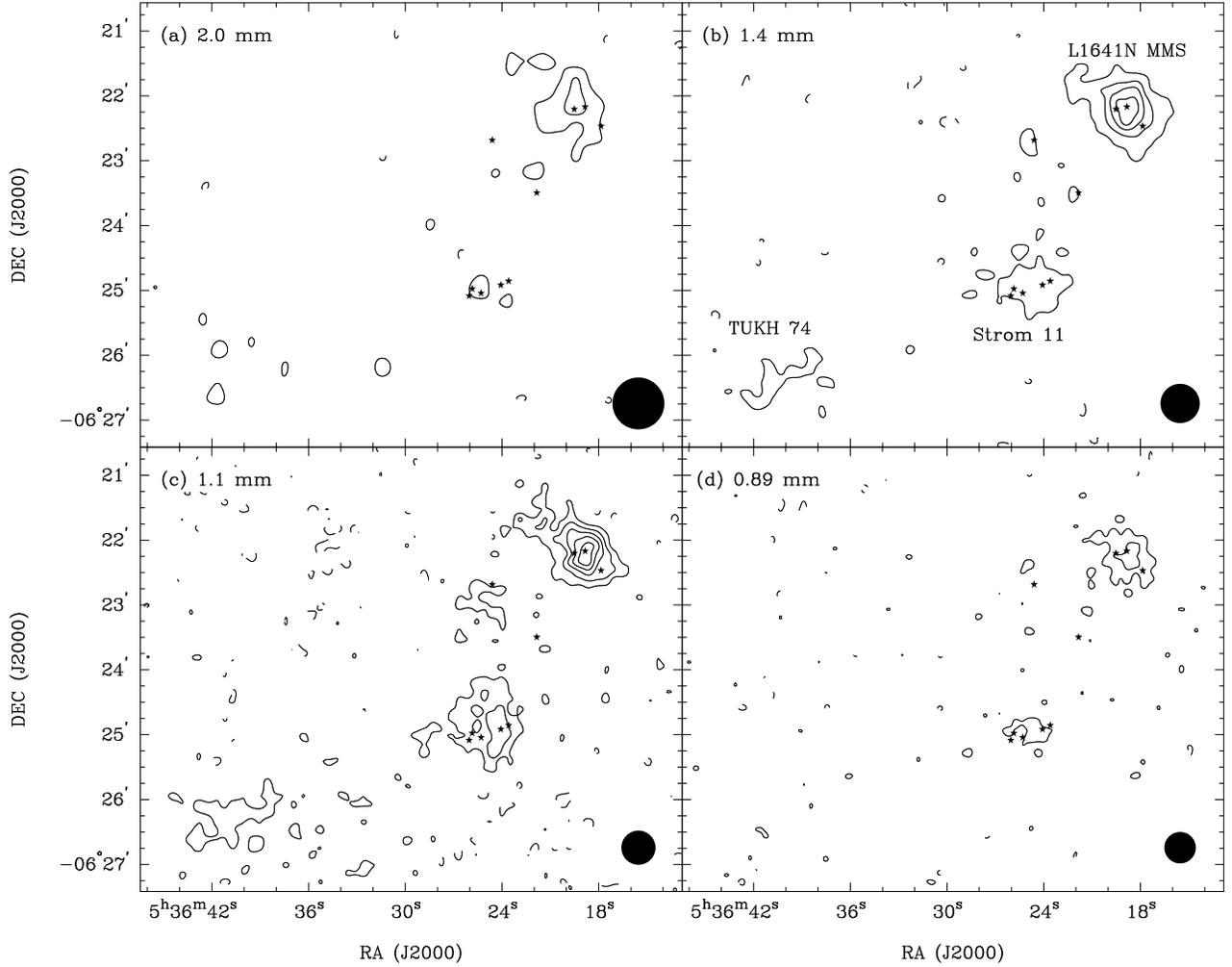}
\caption{
Maps of the L1641N region in the (a) 2.0 mm, (b) 1.4 mm, (c) 1.1 mm,
and (d) 0.89 mm continuum emission.
The contour levels are 1, 2, 3, 4, and 5 $\times$ $3\sigma$,
where $\sigma$ is the noise rms level of each map (Table 2).
Dashed contours are for negative levels.
Shown in the bottom right-hand corner of each panel is the beam size (Table 2).
Detected continuum sources are labeled.
Star symbols:
protostars in the HOPS catalog (Furlan et al. 2016).}
\end{figure*}

\begin{figure*}[!t]
\epsscale{0.9}
\plotone{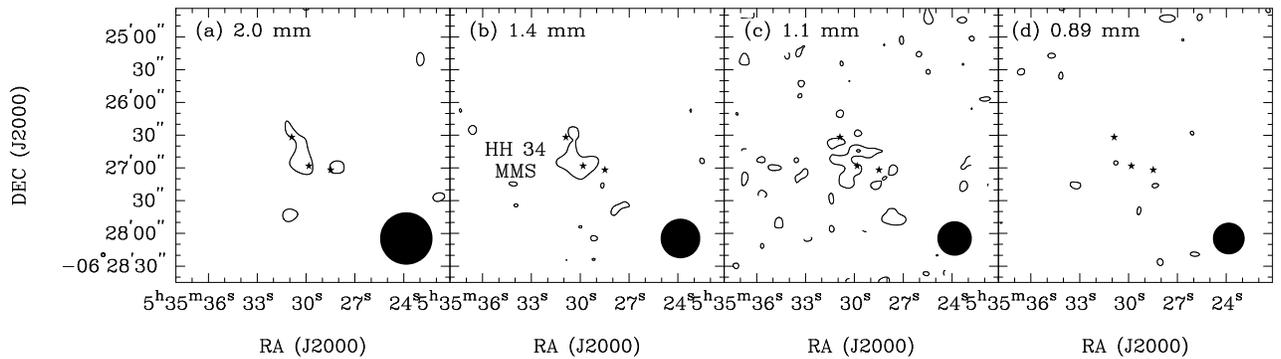}
\caption{
Maps of the HH 34 region in the millimeter--submillimeter continuum emission.
See Figure 10 for the contour levels and markers.}
\end{figure*}

\begin{figure*}[!t]
\epsscale{0.9}
\plotone{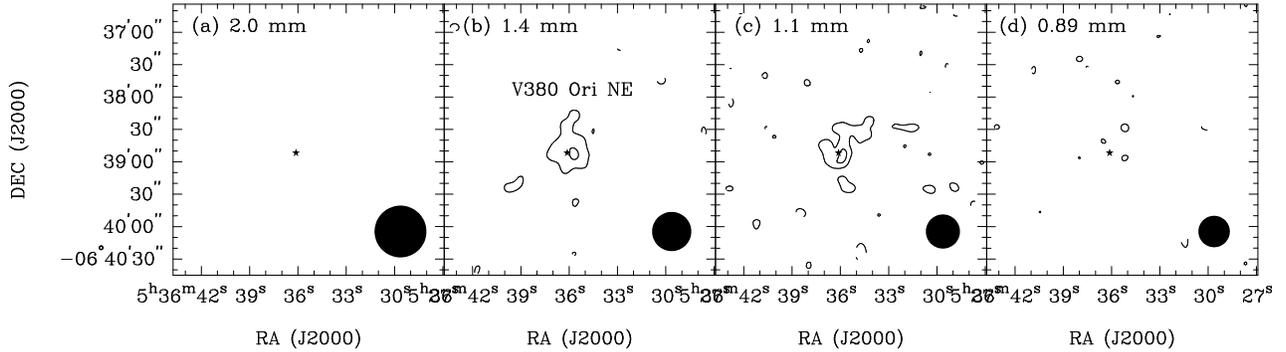}
\caption{
Maps of the V380 Ori NE region
in the millimeter--submillimeter continuum emission.
See Figure 10 for the contour levels and markers.}
\end{figure*}

\begin{figure*}[!t]
\epsscale{0.9}
\plotone{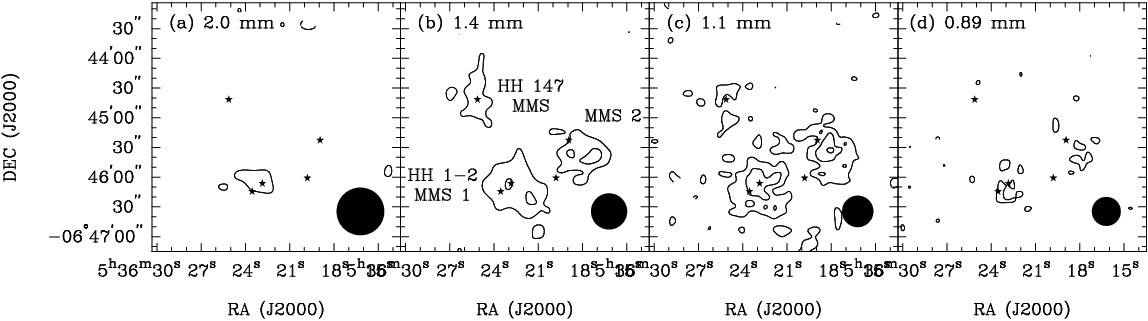}
\caption{
Maps of the HH 1--2 region
in the millimeter--submillimeter continuum emission.
See Figure 10 for the contour levels and markers.}
\end{figure*}

There is a cluster of three cores in the L1641N region,
as shown in Figure 10.
One of them, TUKH 74, is a starless core.
HH 34 MMS and V380 Ori NE are relatively isolated cores (Figures 11--12).
There is another cluster of three cores in the HH 1--2 region (Figure 13).
The star formation activities of individual cores
are summarized in Appendix.
Integrated flux densities were measured for each core.
The flux densities and statistical uncertainties are listed in Table 4.
The uncertainties are dominated by the statistical uncertainties
rather than the calibration uncertainties.

The correction factors for the high-pass filter listed in Table 2
are not exactly correct for extended sources.
The strongest and most extended source in the field of view is L1641N MMS.
Elliptical Gaussian fits to the 1.4 and 1.1 mm maps
give a deconvolved size of FWHM $\approx$ 44$''$ $\times$ 20$''$.
The correction factors appropriate for this source size
are smaller by $\sim$3\%.
As shown in Table 5,
the effect of source size on the flux density measurements
is smaller than the uncertainties.
Moreover, the amount of this effect
is similar across all the wavelength bands,
and the spectral slope does not change significantly.
The other sources in the field are weaker and smaller than L1641N MMS.
Therefore, at the sensitivity level of the MUSIC data,
the correction for source size is unnecessary.

\section{DISCUSSION}

\subsection{V380 Ori NE Precessing Outflow}

The thermal SiO lines are excellent tracers
of the primary jets driven by protostars
because the SiO abundance in the jet
is enhanced by several orders of magnitude
compared to that of the ambient molecular cloud
(Dutrey et al. 1997; Choi 2005; Hirano et al. 2006;
Cabrit et al. 2007; Codella et al. 2007; Choi et al. 2011b).
The SiO spectra of the V380 Ori NE outflow
(Figure 5; also see Figure 6 of Gibb et al. 2004) show
that there is no emission component at the systemic velocity,
which suggests that the SiO line exclusively traces the jet.
Interferometric images of the SiO lines often provide
valuable information on the spatial and velocity structure of the primary jet.
However, the SiO lines are poor tracers of the mass and related quantities
because the abundance varies greatly
(Gibb et al. 2004; L{\'o}pez-Sepulcre et al. 2016).
In this paper, we concentrate on the analysis
of the morphology and kinematics of the SiO jet.

To explain the curved morphology of the H$_2$ jet,
Davis et al. (2000) suggested
that both northern and southern jets are deflected by dense ambient gas
at roughly the same distance from the protostar.
While the H$_2$ image shows only the brightest part of the jet,
the SiO image shows a more comprehensive picture
and seems to rule out the deflection model.
A good example of a deflected jet is
the asymmetric bipolar jet of NGC 1333 IRAS 4A (Choi 2005).
The northeastern jet of IRAS 4A shows a sharp bend,
but the southwestern jet does not.
In the case of V380 Ori NE,
the point-symmetric patterns of the jet morphology and kinematics
suggest that the outflow variability is intrinsic to the driving source.
Moreover, the SiO jet direction and velocity oscillate
along smooth curves and without any abrupt change,
which also supports the intrinsic variability
and rules out external perturbation.
Among the intrinsic variability mechanisms proposed
(Eisl{\"o}ffel \& Mundt 1997; Fendt \& Zinnecker 1998),
the precession of a jet axis seems to be the only viable one.
The orbital motion of a jet source in a binary system
can produce an oscillation pattern,
but the jet morphology would show a mirror symmetry
(see Figure 2 of Fendt \& Zinnecker 1998).
The Lorentz forces caused by external magnetic fields
can produce a point-symmetric bending,
but the jet would not show oscillations.

To measure quantities describing the variability,
the position angle and velocity of the outflow peaks are plotted
as functions of the angular distance from the driving source (Figure 6),
and they were fitted with sinusoidal functions.
The number of free parameters can be reduced
by assuming a perfect point symmetry.
That is, the northern and southern jets
were assumed to have the same oscillation period and amplitudes.
The position angle and line-of-sight velocity
depend on the inclination angle of the velocity vector at each flow element
in a way that their oscillation patterns
have a phase difference of $\pi$/2.
The fitting functions are
\begin{equation}
   p = A_p \cos\left({{2\pi d}\over{L}}+\varphi_0\right) + D_p
\end{equation}
\begin{equation}
   v_p = A_v \sin\left({{2\pi d}\over{L}}+\varphi_0\right) + D_v,
\end{equation}
where $p$ is the position angle (minus 180$^\circ$ for the southern jet)
with respect to the driving source,
$v_p$ is the centroid velocity ($V_c$) with respect to the systemic velocity
($V_c - V_0$ for the southern jet and $V_0 - V_c$ for the northern jet),
$A_p$ and $A_v$ are the oscillation amplitudes,
$d$ is the angular distance from the driving source,
$L$ is the wavelength of oscillation,
$\varphi_0$ is the phase of oscillation at the center,
and $D_p$ and $D_v$ are the mean values.
Fitting all the data points together,
the best-fit values are
$L$ = 28$''$ $\pm$ 2$''$, $\varphi_0$ = 5.0 $\pm$ 0.6,
$A_p$ = 3\fdg5 $\pm$ 0\fdg5, $D_p$ = --8\fdg5 $\pm$ 0\fdg5,
$A_v$ = 2.0 $\pm$ 0.5 km s$^{-1}$, and $D_v$ = 8.2 $\pm$ 0.3 km s$^{-1}$.

\begin{deluxetable*}{llllcccrcccc}
\tabletypesize{\small}
\tablecaption{Detected L1641 Continuum Sources}%
\tablewidth{0pt}
\tablehead{
\colhead{Source} & \colhead{Name}
 & \multicolumn{2}{c}{Peak Position\tablenotemark{a}}
&& \multicolumn{3}{c}{Flux Density\tablenotemark{b}}
&& \multicolumn{3}{c}{Associated Objects\tablenotemark{c}} \\
\cline{3-4} \cline{6-8} \cline{10-12}
&& \colhead{$\alpha_{\rm J2000}$} & \colhead{$\delta_{\rm J2000}$}
&& \colhead{1.4 mm} & \colhead{1.1 mm} & \colhead{S/N}
&& \colhead{OrionAS} & \colhead{HOPS} & \colhead{TUKH}}%
\startdata
1 & L1641N MMS    & 05 36 18.8 & --06 22 19 && 1.52 & 3.4 & 18.3
                  && 536196--62209 & 181--183      & 67 \\
2 & Strom 11      & 05 36 25.5 & --06 24 52 && 0.63 & 1.7 &  7.1
                  && 536248--62454 & 173--176, 380 & \ldots \\
3 & TUKH 74       & 05 36 41.0 & --06 26 16 && 0.50 & 1.1 &  5.7
                  && 536414--62627 & \ldots        & 74 \\
4 & HH 34 MMS     & 05 35 30.7 & --06 26 53 && 0.58 & 1.3 &  5.8
                  && 535297--62701 & 188--190      & 71 \\
5 & V380 Ori NE   & 05 36 35.8 & --06 38 54 && 0.76 & 1.4 &  8.3
                  && 536361--63857 & 169 (362)     & 89 \\
6 & HH 147 MMS    & 05 36 25.5 & --06 44 39 && 0.49 & 1.0 &  4.8
                  && 536259--64440 & 166           & 91 \\
7 & HH 1--2 MMS 2 & 05 36 18.6 & --06 45 39 && 0.74 & 1.7 &  8.4
                  && 536179--64544 & 167, 168      & \ldots \\
8 & HH 1--2 MMS 1 & 05 36 23.1 & --06 46 09 && 0.70 & 1.7 &  8.5
                  && 536229--64618 & 165, 203      & 92 \\
\enddata
\tablenotetext{a}{Units of right ascension are hours, minutes, and seconds,
                  and units of declination are degrees, arcminutes,
                  and arcseconds.}
\tablenotetext{b}{Peak flux densities in Jy beam$^{-1}$,
                  and signal-to-noise ratio in the combined map
                  of the 1.4 and 1.1 mm bands.}%
\tablenotetext{c}{OrionAS: submillimeter continuum sources
                  in Nutter \& Ward-Thompson (2007).
                  HOPS: protostars in Furlan et al.  (2016).
                  TUKH: CS cores in Tatematsu et al. (1993).}%
\end{deluxetable*}

The first two emission peaks (b1 and r1)
have relatively large uncertainties,
and there is a large velocity jump between peaks r1 and r2.
The jump probably indicates
that the jet may be still accelerating in this area.
To understand whether these first peaks
affect the determination of the oscillation parameters,
the fit was performed again without these data points.
The best-fit values are then
$L$ = 28$''$, $\varphi_0$ = 5.1, $A_p$ = 3\fdg5, $D_p$ = --8\fdg6,
$A_v$ = 2.0 km s$^{-1}$, and $D_v$ = 8.3 km s$^{-1}$.
The differences from the values in the previous paragraph
are smaller than the corresponding uncertainties.
Therefore, the derived oscillation parameters are robust.

The oscillation parameters can be related
to the parameters describing the precessing jet.
Assuming that the SiO outflow peaks are confined
to the surface of a circular cone,
\begin{equation}
   \tan A_p = \tan \theta_p / \cos i_p,
\end{equation}
where $\theta_p$ is the half-angle of the precession cone opening,
and $i_p$ is the inclination angle of the precession cone axis
with respect to the plane of the sky.
Assuming that each flow element is moving ballistically
at a uniform speed and along a straight line away from the driving source,
\begin{equation}
   A_v = v_f \cos i_p \sin \theta_p
\end{equation}
\begin{equation}
   D_v = v_f \sin i_p \cos \theta_p,
\end{equation}
where $v_f$ is the flow speed.
Solving these equations,
$i_p$ = 13\fdg5 $\pm$ 3\fdg1, $\theta_p$ = 3\fdg4 $\pm$ 0\fdg5,
and $v_f$ = 35 $\pm$ 8 km s$^{-1}$.
The precession cone is close to the plane of the sky,
the jet-driving disk may be viewed nearly edge-on,
and the velocity vectors of flow elements
have inclinations ranging from 10$^\circ$ to 17$^\circ$.

The coverage of the SiO jet in this paper
corresponds to $\sim$1.3 precession cycles.
To call the variability strictly ``periodic'',
a significantly longer coverage is necessary.
The large-scale structure of the precessing jet (Section 4.2)
may be more complicated than the simple description given here,
but the jet should be covered over two cycles or more
to uncover more details.
Future observations over a larger region,
by way of either deeper integration in the SiO line
or imaging with a different tracer such as CH$_3$OH,
may be helpful.

Figure 6(a) shows that there is a small but noticeable deviation
from the perfect point symmetry.
The data points of northern jet mostly lies above those of southern jet.
The difference is $\sim$4$^\circ$.
Because this deviation is clearer at larger distances,
the position uncertainty of the protostar is not the cause.
One possible explanation is a westward motion of the protostar
relative to the ambient cloud.

\subsection{Timescales and Evolutionary Status of V380 Ori NE}

The parameters of the precessing jet provide interesting information
on the evolutionary status of the V380 Ori NE system.
Assuming that the distance to V380 Ori NE
is the same as that of Orion KL (418 $\pm$ 6 pc; Kim et al. 2008),
the oscillation wavelength corresponds to 1.2 $\times$ 10$^4$ au,
and the precession period is 1600 $\pm$ 400 years.
The length of CO outflow is
$\sim$120$''$ in the $J$ = 1 $\rightarrow$ 0 line (Section 3.2)
and $\sim$100$''$ in the $J$ = 4 $\rightarrow$ 3 line
(Davis et al. 2000),
and there is no sign of outflow on a larger scale.
Assuming that the length of the primary jet
is the same as that of the CO outflow,
the dynamical timescale of the V380 Ori NE outflow system
is $\sim$6300 years.
This timescale may be a good estimate of the age of the protostar.
In general, the outflow timescale has often been used to infer the age,
but unreliable inclination estimates are a major source of ambiguity
(Downes \& Cabrit 2007).
In the case of the SiO outflow of V380 Ori NE,
the combination of oscillation patterns of position angle and velocity
allows an accurate estimation of the inclination angle,
which resolves the ambiguity.

Rough estimates of protostellar mass can be made
from the timescale and flow speed.
Assuming a steady accretion with a ``typical'' accretion rate
of 2 $\times$ 10$^{-6}$ $M_\odot$ yr$^{-1}$ (Shu et al. 1987),
the growth in mass during the outflow timescale is $\sim$0.013 $M_\odot$.
If the protostar initially grew in a first hydrostatic core,
the accretion rate would have been much higher (Saigo et al. 2008).
This high-accretion phase may have lasted only for a relatively short time
until the mass of the first core was exhausted,
which was followed by the normal accretion phase.
Assuming that the mass of the first core
was $\sim$0.01 $M_\odot$ (Saigo et al. 2008),
the current mass of the protostar may be $\sim$0.02 $M_\odot$.
Alternatively, the outflow speed may be comparable to the Kepler speed
on the surface of the protostar (Machida et al. 2008).
Assuming that the protostar radius is in the range of 1--4 $R_\odot$
(Machida et al. 2008; Masunaga \& Inutsuka 2000),
a Kepler speed of 35 km s$^{-1}$ suggests
a stellar mass in the range of 0.006--0.03 $M_\odot$.
Therefore, 0.02 $M_\odot$ may be
a reasonable estimate of the protostellar mass.

The accretion luminosity can be calculated
using the mass, accretion rate, and stellar radius in the previous paragraph.
The expected luminosity is 0.3--1.2 $L_\odot$.
The luminosity of the protostar itself
(from contraction and deuterium burning) can be ignored
because the age is too young (Myers et al. 1998).
By contrast, infrared observations of V380 Ori NE
indicate $L_{\rm bol}$ = 3.9 $L_\odot$ (Furlan et al. 2016).
Considering the uncertainties and assumptions involved,
it is unclear whether or not the difference
between the expected and observed luminosities is significant.
If it is, then this discrepancy suggests several possibilities:
(1) the protostar may be more massive;
(2) the system may be currently in a state of enhanced accretion;
(3) there could be multiple protostars;
(4) the bolometric luminosity may be overestimated,
    considering that the infrared flux contains
    a contribution from the bright outflow knot.
We use the protostellar mass of 0.02 $M_\odot$ in the discussion below,
but note that there is a possibility of larger mass.

The V380 Ori NE system is still extremely young,
and the protostar may be evolving significantly over a jet precession period,
which corresponds to about a quarter of the outflow timescale.
For example, the mass may increase by more than 10\%
during a precession cycle.
If the jet-driving protostar belongs to a binary system (Section 4.3),
the matter accreting from the envelope may distribute
the accompanying angular momentum
into protostellar spins, disk rotation, and binary orbital motions.
The quantities related to the jet precession,
such as flow speed and precession period,
then may also evolve at significant rates,
and the oscillation pattern of the jet may be more complicated
than the simple sinusoidal profiles described in Section 4.1.
Indeed, the CO $J$ = 4 $\rightarrow$ 3 line map shows
a change of outflow direction on a large scale (Davis et al. 2000).
Davis et al. (2000) inferred
an oscillation wavelength of $\sim$200$''$ for the CO outflow.
If the CO and SiO outflows are tracing the same precession phenomenon,
either the precession rate may be increasing
or the flow speed may be decreasing.
Alternatively, the outflow variability may be indicating
a superposition of multiple precession modes.

\subsection{Precession Mechanism and Binary System}

The point-symmetric and oscillating pattern of the V380 Ori NE jet suggests
that the jet axis (i.e., the rotation axis of jet-launching mechanism)
is precessing.
The usual explanation for jet precession is
that the jet-driving protostar belongs to a non-coplanar binary system,
i.e., the jet-launching disk is misaligned with the binary orbital plane
(Terquem et al. 1999; Bate et al. 2000; Sheikhnezami \& Fendt 2015).
The disk responds to the tidal perturbation mainly in two different modes:
precession (response to the axially symmetric potential)
and wobbling (relatively faster response
to the azimuthal Fourier component of second harmonic).

\begin{deluxetable*}{lccccccr}
\tabletypesize{\small}
\tablecaption{L1641 Continuum Source Parameters}%
\tablewidth{0pt}
\tablehead{
\colhead{Name} & \colhead{$R_A$\tablenotemark{a}}
& \multicolumn{4}{c}{Flux Density\tablenotemark{b}}
& \colhead{$\beta$\tablenotemark{c}} & \colhead{Mass\tablenotemark{c}} \\
\cline{3-6}
& & \colhead{2.0 mm} & \colhead{1.4 mm} & \colhead{1.1 mm} & \colhead{0.89 mm}
& & \colhead{($M_\odot$)}}%
\startdata
L1641N MMS\tablenotemark{d}
              & 60$''$ & 1.07 $\pm$ 0.20 & 3.04 $\pm$ 0.52
              & 6.4 $\pm$ 1.2 &    19.9 $\pm$ 6.2
              & 1.4 $\pm$ 0.3 & 11.6 $_{-3.9}^{+5.8}$ \\
Strom 11      & 48$''$ & 0.31 $\pm$ 0.10 & 1.34 $\pm$ 0.21
              & 4.1 $\pm$ 0.6 &    11.1 $\pm$ 2.2
              & 2.2 $\pm$ 0.3 & 16.9 $_{-6.1}^{+9.5}$ \\
TUKH 74       & 51$''$ & 0.42 $\pm$ 0.10 & 1.12 $\pm$ 0.29
              & 3.0 $\pm$ 0.4 & \phn5.7 $\pm$ 1.5
              & 1.4 $\pm$ 0.3 &  4.3 $_{-1.5}^{+2.3}$ \\
HH 34 MMS     & 48$''$ & 0.41 $\pm$ 0.12 & 0.73 $\pm$ 0.11
              & 1.9 $\pm$ 0.4 & \ldots           
              & 0.7 $\pm$ 0.6 &  1.0 $_{-0.6}^{+1.4}$ \\
V380 Ori NE   & 36$''$ & \ldots          & 0.71 $\pm$ 0.11
              & 1.4 $\pm$ 0.2 & \ldots           
              & 0.3 $\pm$ 0.5 &  0.4 $_{-0.2}^{+0.4}$ \\
HH 147 MMS    & 40$''$ & \ldots          & 0.67 $\pm$ 0.19
              & 1.2 $\pm$ 0.3 & \ldots           
              & 0.7 $\pm$ 0.7 &  0.8 $_{-0.5}^{+1.1}$ \\
HH 1--2 MMS 2 & 38$''$ & \ldots          & 0.95 $\pm$ 0.11
              & 2.7 $\pm$ 0.3 & \phn6.0 $\pm$ 1.0
              & 1.9 $\pm$ 0.4 &  6.7 $_{-2.6}^{+4.1}$ \\
HH 1--2 MMS 1 & 48$''$ & 0.43 $\pm$ 0.11 & 1.30 $\pm$ 0.23
              & 3.7 $\pm$ 0.6 & \phn7.9 $\pm$ 2.0
              & 1.6 $\pm$ 0.3 &  6.6 $_{-2.3}^{+3.6}$ \\
\enddata
\tablenotetext{a}{Radius of the circular aperture for flux measurements.}
\tablenotetext{b}{Integrated flux densities in Jy.}%
\tablenotetext{c}{Dust emissivity index and mass of molecular gas,
                  assuming $T_d$ = 20 K.}%
\tablenotetext{d}{Parameters from Table 5 method E + A.}
\end{deluxetable*}

\begin{deluxetable*}{lccccc}
\tabletypesize{\small}
\tablecaption{L1641N MMS Photometry}%
\tablewidth{0pt}
\tablehead{
\colhead{Method\tablenotemark{a}}
& \multicolumn{4}{c}{Flux Density\tablenotemark{b}} & \colhead{$\beta$} \\
\cline{2-5}
& \colhead{2.0 mm} & \colhead{1.4 mm} & \colhead{1.1 mm} & \colhead{0.89 mm}}%
\startdata
U + A & 1.05 $\pm$ 0.20 & 2.94 $\pm$ 0.50 & 6.2 $\pm$ 1.1 & 19.0 $\pm$ 5.9
      & 1.4 $\pm$ 0.3 \\
E + A & 1.07 $\pm$ 0.20 & 3.04 $\pm$ 0.52 & 6.4 $\pm$ 1.2 & 19.9 $\pm$ 6.2
      & 1.4 $\pm$ 0.3 \\
E + G & 1.17 $\pm$ 0.26 & 2.82 $\pm$ 0.39 & 6.6 $\pm$ 1.1 & 18.1 $\pm$ 3.1
      & 1.6 $\pm$ 0.3 \\
\enddata
\tablenotetext{a}{U: The correction for the effect of high-pass filter
                     was made using $\eta_{\rm hpf}$ for an unresolved source,
                     listed in Table 2.
                  E: The correction was made
                     using $\eta_{\rm hpf}$ = 0.87, 0.89, 0.90, and 0.90,
                     which is appropriate for the size of L1641N MMS.
                  A: The flux density was integrated in a circular aperture.
                  G: The peak flux density was measured by fitting the image
                     with an elliptical Gaussian profile,
                     and scaled for the solid angle of the source.}
\tablenotetext{b}{Integrated flux densities in Jy.}%
\end{deluxetable*}

First, the disk-jet system as a whole can precess.
The precession period may be longer than the binary orbital period
by an order of magnitude.
Bate et al. (2000) suggested a factor of $\sim$20
for a typical protostellar binary system.
If the disk precesses for a period of 1600 years,
the inferred orbital period is $\sim$80 years.
The separation between the binary members would then be $\sim$6 au
(0\farcs014 at 418 pc).
Because the disk size should be smaller than the binary orbit,
the disk radius cannot be larger than $\sim$2 au.

Second, the disk axis can wobble because of the tidal force,
and the jet axis can precess
along with the wobbling of the inner part of the disk (Bate et al. 2000).
The period of wobble may be half the binary orbital period.
From the jet precession period of 1600 years,
the orbital period would be $\sim$3200 years.
The binary separation would be $\sim$70 au (0\farcs17 at 418 pc),
and the disk radius can be up to $\sim$20 au.

The inferred binary separation provides
an interesting insight into the binary formation process.
In general, there are two channels for binary formation:
fragmentation of the disk by gravitational instability
and turbulent fragmentation of the dense cloud core.
Theoretical models and observational imaging of binary systems suggest
that the disk instability and the turbulent fragmentation processes
produce systems with binary separations
smaller than $\sim$100 au and larger than $\sim$500 au, respectively
(Offner et al. 2010; Kratter \& Lodato 2016;
Tobin et al. 2016; Lee et al. 2017).
A survey of binary separations showed
a bimodal distribution with peaks at $\sim$75 au and $\sim$3000 au,
probably representing the disk instability and turbulent fragmentation modes
of binary formation (Tobin et al. 2016). 
The binary separation of V380 Ori NE may be 6 or 70 au,
depending on the precession mechanism.
Either way, the separation suggests
that this binary system was probably formed
through the disk instability process.

\begin{figure*}[!t]
\epsscale{1.0}
\plotone{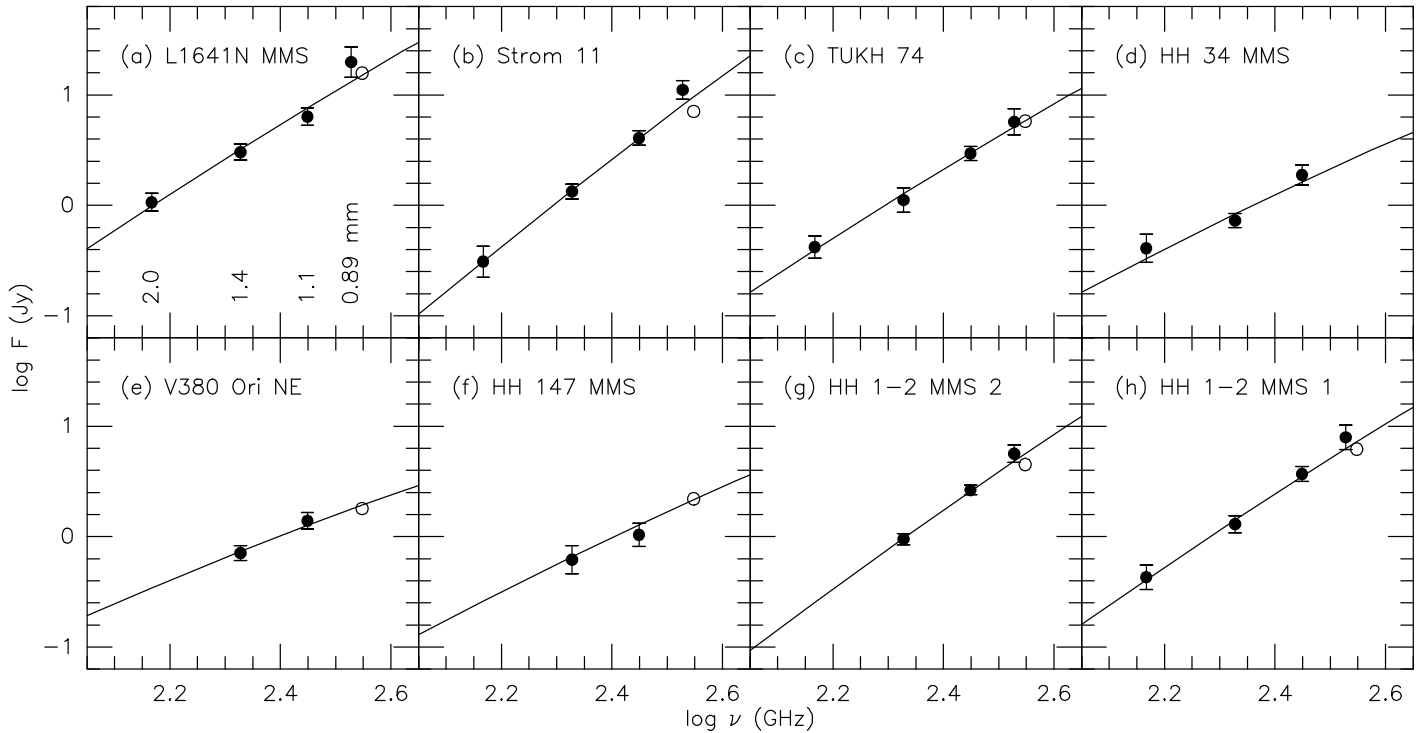}
\caption{
Spectral energy distributions of the detected dense cores.
Filled circles:
flux densities from the MUSIC maps.
Flux uncertainties (1$\sigma$) are shown.
Open circles:
850 $\mu$m flux densities from Nutter \& Ward-Thompson (2007).
Solid curves:
best-fit modified black body spectra, assuming $T_d$ = 20 K.}
\end{figure*}

There are alternative mechanisms of jet precession.
For example, magnetically driven disk warping
can cause the jet to precess (Lai 2003).
In this mechanism,
the disk is assumed to be perturbed by large-scale magnetic fields,
and the protostar does not have to belong to a binary system.

Because the observed length of the SiO jet is
only a little longer than a full precession cycle,
the directional variability can be more complicated
than the simple periodic precession.
In such a case, alternative variability mechanisms are possible.
For example, accretion of material from a turbulent cloud core
can make the disk rotation axis change orientation chaotically
(Bate et al. 2010; Tsukamoto \& Machida 2013).
As a result, the jet axis can wander stochastically (Spalding et al. 2014).
This turbulence-driven precession mechanism may be relevant to V380 Ori NE
because the system is very young.

\subsection{Continuum Spectra of the L1641 Dense Cores}

SEDs of the detected cores are shown in Figure 14.
Because the four wavelength bands were observed simultaneously,
and then calibrated, imaged, and measured consistently,
the spectral slopes based on the MUSIC data set are more reliable
than those derived from inhomogeneous data sets.
The detected cores span a wide range of spectral slope.
When the flux density, $F_\nu$, is expressed
as a power-law function of the frequency, $\nu$,
the spectral index, $\alpha$ (i.e., $F_\nu \propto \nu^\alpha$),
ranges from 2.0 to 3.9.
The measured spectral slopes are
consistent with that of thermal dust emission.

Before discussing the dust continuum emission further,
it is necessary to consider possible contributions
from other components of radiation.
In particular, V380 Ori NE, HH 34 MMS, and HH 147 MMS
show spectral slopes relatively shallower than the others,
and ``contaminations'' in the long wavelength bands need to be examined.
First, molecular lines in the continuum bands can affect the spectral slope.
Practically, only very bright CO lines or maser lines
can cause noticeable effects.
The MUSIC 1.4 and 0.89 mm bands include
the CO $J$ = 2 $\rightarrow$ 1 and 3 $\rightarrow$ 2 lines, respectively.
Even if these bands contain extra flux densities from the CO lines,
they would not necessarily make the slope shallower.
The contamination by CO, especially the $J$ = 3 $\rightarrow$ 2 line,
would steepen the spectral slope
and cannot be responsible for the shallow slopes.
Moreover, the three shallow-spectrum cores
were undetected in the 0.89 mm image,
suggesting that the CO $J$ = 3 $\rightarrow$ 2 line
is too weak to affect the spectral slope.
There is no strong maser line in the MUSIC bands.
Therefore, molecular lines are unlikely to affect the continuum spectra.
Second, some emission components can affect the long wavelength bands.
The free-free emission may have a flat spectrum in the millimeter range,
and the anomalous microwave emission may have a spectrum
peaking at $\sim$25 GHz and quickly decreasing at higher frequencies.
The 3.6 cm flux density of the HH 34 radio jet is 0.16 mJy
(Rodr{\'\i}guez \& Reipurth 1996),
V380 Ori NE was undetected at 6 cm ($<$0.1 mJy; Morgan et al. 1990),
and HH 147 MMS was undetected at 6 cm or 3.6 cm
($<$0.03 mJy; Rodr{\'\i}guez et al. 2000).
Therefore, the long-wavelength emission components
are negligible in the millimeter range.
Finally, external heating by the interstellar radiation field
can increase the flux densities in the long wavelength bands,
though the effect on the spectral slope may be small.
This effect may be noticeable
only for low-luminosity objects ($<$1 $L_\odot$)
in extremely strong radiation fields
(Shirley et al. 2002; Furlan et al. 2016).
Because the bolometric luminosities of HH 34 MMS, V380 Ori NE, and HH 147 MMS
are $L_{\rm bol}$ = 18.8, 3.9, and 15.5 $L_\odot$,
respectively (Furlan et al. 2016),
it is unlikely that the external heating can affect the spectral slopes.
Therefore, the millimeter--submillimeter flux densities of the detected cores
are dominated by thermal dust emission
(see further discussions in the last paragraph of Section 4.5).

\begin{figure*}[!t]
\epsscale{1.0}
\plotone{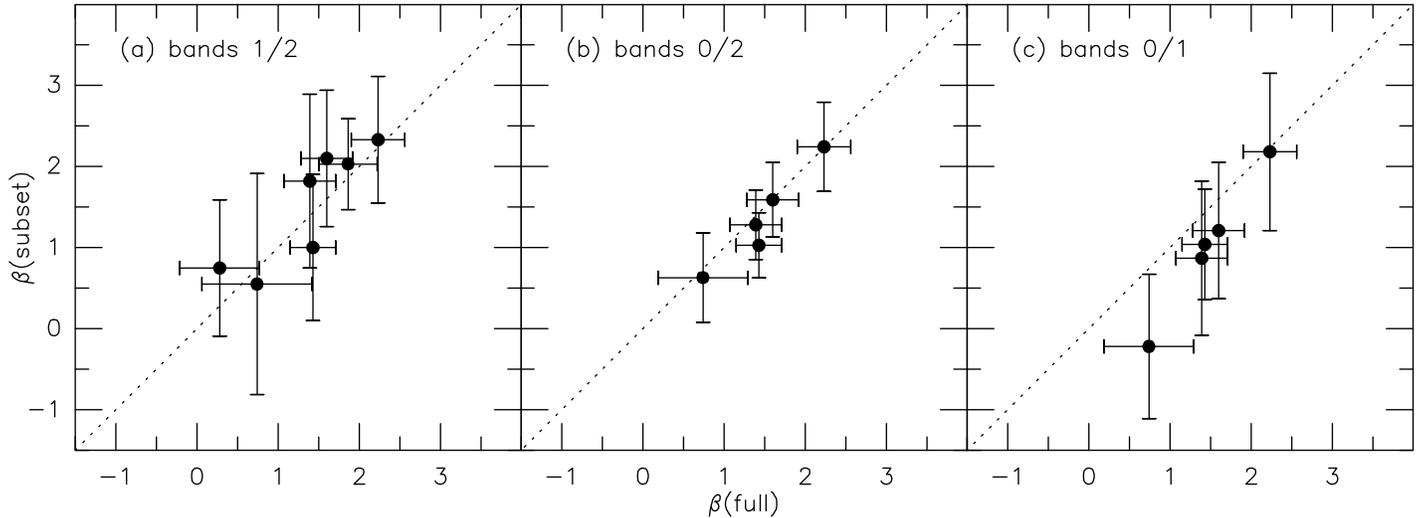}
\caption{
Correlation diagrams of the dust emissivity indices
derived from the full data and from subsets of the MUSIC data:
(a) 1.4 and 1.1 mm bands,
(b) 2.0 and 1.1 mm bands,
and (c) 2.0 and 1.4 mm bands.
HH 34 MMS is missing from panel (a)
because the 850 $\mu$m flux density is not used
for the calculation of $\beta$(full).
V380 Ori NE, HH 147 MMS, and HH 1--2 MMS 2 are missing from panels (b) and (c)
because they were undetected in the 2.0 mm band.}
\end{figure*}

Most of the dust continuum emission in the MUSIC beams
may come from the dense cores,
and the accretion disks around embedded protostars
may contribute only a little.
Interferometric observations with sub-arcseconds resolutions
are needed to measure the flux densities from the disks,
but there is no such observation available for the detected cores.
Stanke \& Williams (2007) observed L1641N MMS
with a $\sim$1\farcs4 resolution,
and detected a compact source, L1641N MM1, with 0.54 Jy at 1.3 mm.
They interpreted this source as an envelope structure rather than a disk.
Thus, the contribution of the disk to the MUSIC flux density
must be much smaller than $\sim$10\%.
Even so, the disk in L1641N MM1 is presumably one of the brightest examples
in the observed region.
Therefore, the contribution from the protostellar disks in the detected cores
is unimportant in the interpretation of the SEDs from MUSIC data.

\subsection{Dust Emissivity and Mass of Dense Cores}

The dust continuum radiation is expected to be optically thin
in the wavelength range covered.
The mass emissivity is often expressed
in a power-law function of the frequency
\begin{equation}
   \kappa_\nu = 0.1 \left({\nu\over{\nu_0}}\right)^\beta
                {\rm cm^2\ g^{-1}},
\end{equation}
where $\nu_0$ = 1 THz, and $\beta$ is the emissivity index
(Hildebrand 1983; Henning et al. 1995;
Andr{\'e} et al. 2010; Sadavoy et al. 2016).
This equation implicitly assumes a gas-to-dust mass ratio of $\sim$100.
The emissivity in the millimeter range is extrapolated from 1 THz
because the emissivity is known better in the infrared range
(Henning et al. 1995).

The SED can be fitted with a modified blackbody function,
\begin{equation}
   F_\nu = {M\over{D^2}} \kappa_\nu B_\nu(T_d),
\end{equation}
where $M$ is the mass of molecular gas, $D$ is the distance to the source,
$B_\nu$ is the Planck function, and $T_d$ is the dust temperature.
In the millimeter wavelength range,
the SED fit is almost insensitive to $T_d$,
and $\beta$ depends on $T_d$ only weakly.
A typical temperature for dense cores is $T_d$ = 20 K
(Johnstone \& Bally 2006).
Specifically, Stutz et al. (2013) derived 20.4 K for V380 Ori NE
by fitting the SED in the 70--870 $\mu$m range.

SED fits to the measured flux densities were made assuming $T_d$ = 20 K.
Four cores were detected at 0.89 mm,
and their flux densities are consistent
with the 850 $\mu$m flux densities
reported by Nutter \& Ward-Thompson (2007).
Therefore, the MUSIC data and the 850 $\mu$m data
were used together for the SED fits.
For the purpose of weighting,
the uncertainty of the 850 $\mu$m data were set to be 20\%,
which gives weights roughly similar to those of the MUSIC data points.
The 850 $\mu$m flux density was not used for HH 34 MMS
because there is a secondary source (IRS 5) near the edge of the main source,
which makes it difficult to treat the data consistently.

Figure 14 shows the best fits, and Table 4 lists the derived $\beta$.
The single-component fits are reasonably good,
and most data points agree with the fits within the uncertainties.
The average $\beta$ is 1.3,
which is in the range of typical $\beta$ values of dense cores (Section 4.6).
V380 Ori NE, HH 34 MMS, and HH 147 MMS
have particularly lower $\beta$ values
(0.3, 0.7, and 0.7, respectively) than the other cores,
suggesting that they contain a considerable amount of large dust grains.
Masses of the cores were derived from the SED fits (Table 4),
assuming $D$ = 418 pc (Kim et al. 2008).
The core masses in the literature are often calculated
using a fixed value of emissivity at a certain frequency.
For example, Nutter \& Ward-Thompson (2007)
assumed 0.01 cm$^2$ g$^{-1}$ at 850 $\mu$m,
which corresponds to $\beta$ $\approx$ 2.2 in Equation (6).
For the three cores with relatively high $\beta$
(Strom 11, HH 1--2 MMS 2, and HH 1--2 MMS 1),
the mass estimates in this paper agree
with those of Nutter \& Ward-Thompson (2007) within 40\%.
For the three cores with low $\beta$ (V380 Ori NE, HH 34 MMS, and HH 147 MMS),
the masses in Table 4 are smaller than those of Nutter \& Ward-Thompson (2007)
by a factor of $\sim$4.
Note that the mass estimates given here critically depend
on the emissivity assumed in Equation (6),
and it can be a major source of uncertainty.
For example, the proportionality constant can be different
if the cores have non-standard gas-to-dust mass ratios.
The simple power-law description can also be an issue,
because the index $\beta$ is not necessarily a constant
over a large range of frequency (Draine 2006; Coupeaud et al. 2011).

While the inclusion of the 850 $\mu$m data
reduces the uncertainty in $\beta$,
especially for the sources undetected in the 0.89 mm band,
it makes the data set inhomogeneous.
There is a possibility of bias coming from the differences
in calibration, filtering, and photometry.
A more aggressive filtering of the 850 $\mu$m data
can bias $\beta$ toward lower values.
To understand the effect of the 850 $\mu$m data,
the emissivity index was calculated using the 1.4 and 1.1 mm band data
and compared with that of the full data set.
Figure 15(a) shows that there is no obvious bias.
The range of $\beta$ values and the order among the sources
are largely unchanged.
For V380 Ori NE and HH 147 MMS,
the fits to the 1.4 and 1.1 mm band data
give $\beta$ = 0.7 $\pm$ 0.8 and 0.5 $\pm$ 1.4, respectively.
Though these emissivity indices are still less than one,
the uncertainties are large.
The low $\beta$ values of these cores are, at most, very tentative,
and consequently the discussion given below
(on the pre-protostellar core lifetime) is also tentative.

The emissivity index from the SED fit
has a weak dependency on the dust temperature assumed.
In the Rayleigh-Jeans (high temperature) limit,
the $\beta$ value would be lower by 0.3 than that of $T_d$ = 20 K.
A lower $T_d$ gives a higher $\beta$ value.
Radiative transfer models of protostellar envelope show
that the dust temperature decreases outward from the center,
hits a minimum temperature of $\sim$10 K,
and increases again near the edge of the cloud,
owing to the heating by the interstellar radiation field
(Shirley et al. 2002).
The $\beta$ value from the SED fit assuming $T_d$ = 10 K
can then be considered as an upper limit.
The limits for some of the detected cores are:
$\beta <$ 2.6 for Strom 11, $\beta <$ 1.1 for HH 34 MMS,
and $\beta <$ 0.7 for V380 Ori NE.

To test the presence of any extra emission component,
the emissivity index was calculated using subsets of the MUSIC data
and compared with that of the full data set.
Figure 15(b) shows $\beta$(subset) from the 2.0 and 1.1 mm bands.
These bands do not contain the CO lines.
All the data points are consistent with $\beta$(subset) = $\beta$(full)
within the uncertainties,
which suggests that there is no noticeable effect
of CO lines on the spectral slope.
Figure 15(c) shows the $\beta$(subset) from the 2.0 and 1.4 mm bands.
If there were a significant contribution
from long-wavelength emission components,
such as free-free or anomalous microwave emission,
the corresponding source would show $\beta$(subset) $<$ $\beta$(full).
Most of the data points are located
near the $\beta$(subset) = $\beta$(full) line,
which suggests that there is no significant contamination
from long-wavelength emission components.
The lowest data point (corresponding to HH 34 MMS)
is located marginally below the line.
There are a few possible explanations.
First, the 2.0 mm flux density
may be very uncertain and somehow overestimated.
Second, the two sub-cores (Appendix) may have different dust properties,
and the single-component fit may not work well.
Third, there may be a long-wavelength emission component.
Though the radio jet is not strong enough
to affect the millimeter spectrum (Section 4.4),
a low-level diffuse free-free emission component
(large enough to be resolved out by VLA)
could contaminate the 2.0 mm flux density.
For HH 34 MMS,
the fit to the 1.4 and 1.1 mm band data gives $\beta$ = 1.8 $\pm$ 0.9.
Therefore, the low $\beta$ value of HH 34 MMS is very tentative.

\subsection{Emissivity Index and Other Physical Quantities}

The low $\beta$ values of V380 Ori NE, HH 34 MMS, and HH 147 MMS suggest
that dust grains can grow to a large size in some dense cores.
Microscopically, the size, composition, and structure of dust
can affect emissivity index,
and growth of ice mantles and coagulation of particles
may be the important physical processes in dense cores
(Ossenkopf \& Henning 1994; Henning et al. 1995; Ormel et al. 2009).
Observationally, submillimeter studies of nearby star-forming regions
showed that dense cores usually have $\beta$ = 1.0--2.0
(e.g., Visser et al. 1998; Hogerheijde \& Sandell 2000).
Recent studies using the data in the far-infrared to submillimeter range
showed that dense cores have $\beta$ = 1.0--2.7
(Sadavoy et al. 2016; Chen et al. 2016).
In the diffuse interstellar medium,
molecular-gas dominated regions have $\beta$ $\approx$ 1.66
(Planck Collaboration et al. 2014).
In these studies, however,
the emissivity index may have dependencies on the dust temperature.
Sadavoy et al. (2013) suggested that data in long wavelengths are important.
Schnee et al. (2014) reported $\beta$ = 0.6 for OMC 3 MMS 6,
but their low $\beta$ estimates were later disputed by Sadavoy et al. (2016).
Therefore, the low $\beta$ values of the three cores in this study
are rather unusual.

\begin{figure*}[!t]
\epsscale{1.0}
\plotone{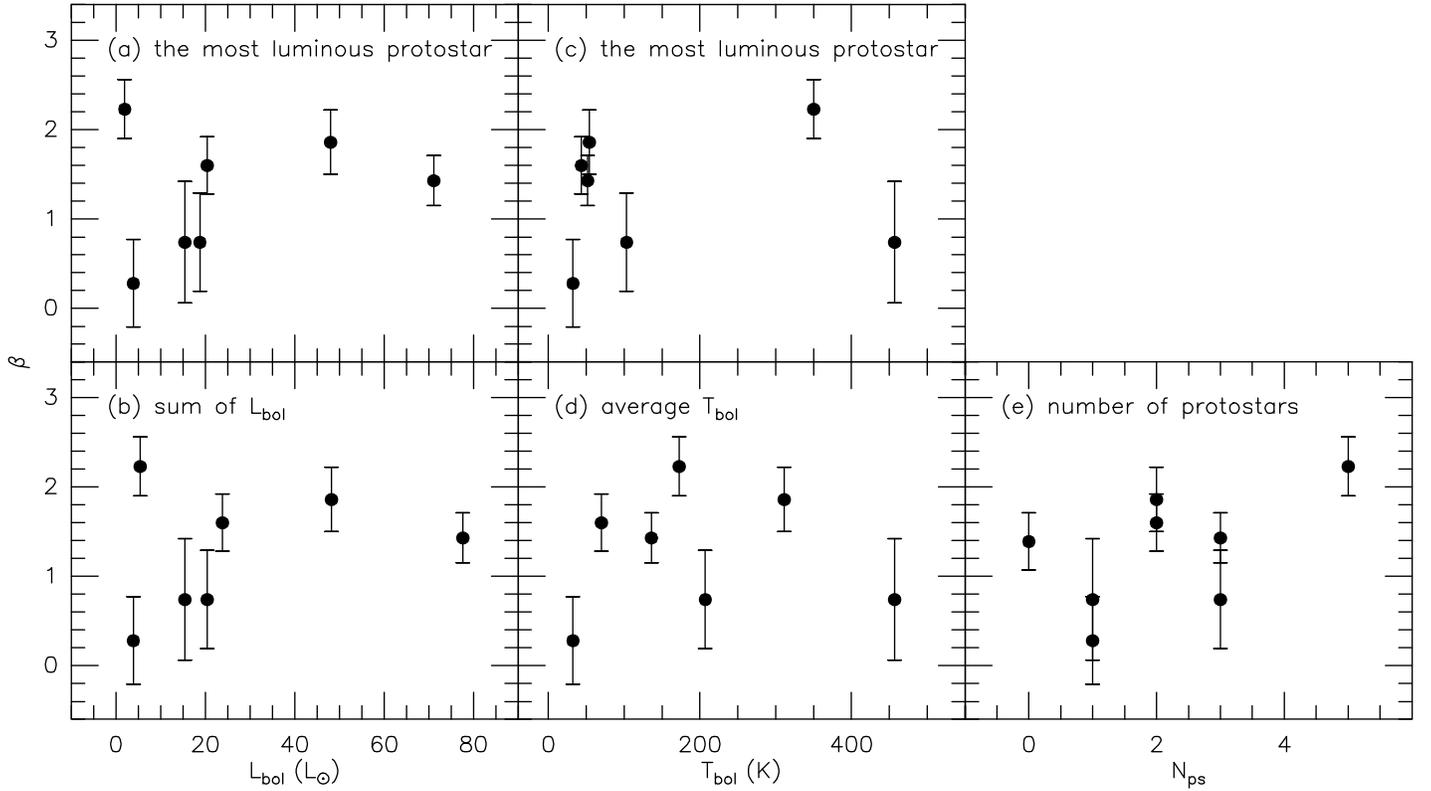}
\caption{
Correlation diagrams of the dust emissivity index
with various properties of the dense cores:
(a) bolometric luminosity of the most luminous protostar,
(b) sum of the bolometric luminosities of the protostars,
(c) bolometric temperature of the most luminous protostar,
(d) average bolometric temperature of the protostars,
and (e) number of protostars in the core.
The bolometric luminosities and temperatures are from Furlan et al. (2016).
TUKH 74 is missing from panels (a--d) because it contains no protostar.}
\end{figure*}

\begin{figure*}[!t]
\epsscale{1.0}
\plotone{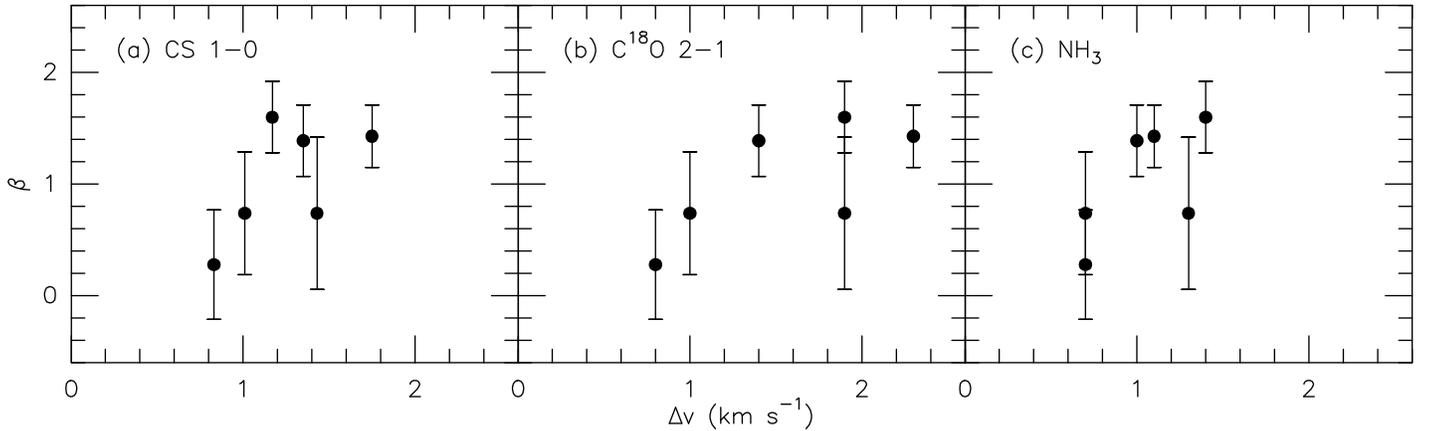}
\caption{
Correlation diagrams of the dust emissivity index with molecular line widths:
(a) CS $J$ = 1 $\rightarrow$ 0 (Tatematsu et al. 1993),
(b) C$^{18}$O $J$ = 2 $\rightarrow$ 1 (Wilson et al. 1999),
and (c) NH$_3$ (1, 1)/(2, 2) (Wilson et al. 1999).
Strom 11 and HH 1--2 MMS 2 are missing from these plots
because there is no counterpart
in the CS core catalog of Tatematsu et al. (1993).}
\end{figure*}

Because the emissivity index spans a wide range,
it would be interesting to examine relations with other physical quantities
that are likely relevant to dust properties.
Figure 16 shows the correlations of the emissivity index
with the properties of protostars embedded in the dense cores.
The multiplicity, $N_{\rm ps}$, in Figure 16(e)
is the number of protostars in the radii given in Table 4.
The correlations with a kinematic property of the molecular gas,
the widths of molecular lines tracing dense gas,
are shown in Figure 17.

Properties of protostars have only weak correlations with $\beta$, if any.
It is unclear if there is a positive correlation
with the bolometric luminosities (Figure 16(a, b)),
and the linear correlation coefficient is $r$ $\approx$ 0.3.
The lack of correlation with bolometric temperatures
($r$ $\approx$ 0.1; Figure 16(c, d)) is surprising
because the variations in dust properties
are often considered to be evolutionary effects
(Visser et al. 1998; Dent et al. 1998).
Froebrich (2005) also found that there is no correlation
between $\beta$ and $L_{\rm bol}$ or $T_{\rm bol}$.
There is a weak correlation with the number of protostars
($r$ = 0.5; Figure 16(e)),
but more data are needed to confirm this correlation.

The strongest correlation with $\beta$
comes from the molecular line widths (Figure 17).
The linear correlation coefficients are $r$ = 0.6, 0.7, and 0.7
for CS, C$^{18}$O, and NH$_3$, respectively.
This correlation suggests
that the turbulence may be a major factor affecting dust properties,
though the exact mechanism is unclear.
The turbulence in dense cores may be
related to the cloud support and fragmentation on large scales
and the motions of dust particles on small scales.
In turn, the turbulence may be affecting the global timescale
and also the local grain size distribution.
It has been suggested that the grain dynamics may be a major factor
in the shattering and coagulation of dust grains
(Chokshi et al. 1993; Hirashita \& Yan 2009; Ormel et al. 2009).
Another relevant process may be the cloud-scale clustering
of dynamically decoupled grains (Hopkins 2014; Hopkins \& Lee 2016). 

\subsection{Timescales and Dust Growth in Dense Cores}

Dust grains can grow to a large size
in the circumstellar structure around relatively evolved YSOs,
such as protoplanetary disks (Testi et al. 2014).
It has been unclear
whether or not dense cores or protostellar envelopes
contain large dust grains,
and if so, when they start to grow such grains.
Kwon et al. (2009) reported that
some Class 0 protostars have $\beta$ values around or smaller than 1,
based on the interferometric observations at 1.3 and 2.7 mm wavelength bands.
They suggested
that the central $\sim$400 au region of the protostellar envelope
contains large grains, up to $\sim$1 cm.
Miotello et al. (2014) suggested
the presence of millimeter-sized dust grains
in the envelopes of Class I protostars.
Large dust grains in protostellar envelopes are interesting
because they can behave like a separate fluid component
and cause intriguing effects,
such as the increase of dust-to-gas ratio in the central region,
spatial segregation of grains by size,
and formation of flattened dust-rich structure
(Bate \& Lor{\'e}n-Aguilar 2017).
This flattened structure can have a radius of several hundreds of au
and is probably observable using radio interferometers.

Ormel et al. (2009) simulated the evolution of dust population in dense cores.
They found that grain growth is negligible in free-fall timescales
($\sim$10$^5$ years for a density of 10$^5$ cm$^{-3}$)
but becomes significant on larger timescales,
such as ambipolar diffusion timescale
($\sim$4 $\times$ 10$^6$ years for a density of 10$^5$ cm$^{-3}$).
Wong et al. (2016) investigated
the dust coagulation process in protostellar envelopes
and found that the grain growth at a density of 10$^5$ cm$^{-3}$
is not fast enough to produce millimeter-sized grains.
They suggested that such grains may be formed
in the central dense (10$^{10}$ cm$^{-3}$ or higher) region of the envelope
and transported outward by the protostellar outflow,
up to a distance of 1000 au in 10$^4$ years.

A $\beta$ value lower than 1 implies
that the source contains dust grains
up to a few times the wavelength observed (Draine 2006),
and the low $\beta$ values of V380 Ori NE, HH 34 MMS, and HH 147 MMS suggest
that these cores, or envelopes of Class 0--I protostars,
contain dust grains as large as several millimeters.
In the porous composite grain model of Ricci et al. (2010),
$\beta$ $\approx$ 0.3 implies a maximum grain size of at least $\sim$7 mm.
If these large grains are mainly icy grains,
$\beta$ $\approx$ 0.3 implies a maximum grain size of $\sim$2 cm or larger
(see Figure 4 of Testi et al. 2014).
For these three cores,
the size of the 850 $\mu$m emission sources
is larger than the $\sim$15$''$ beam (Nutter \& Ward-Thompson 2007).
Therefore, these cores seem to contain millimeter-sized grains
in $\sim$10,000 au scale regions.

The outflow-transportation model of Wong et al. (2016)
cannot explain the presence of the large grains
in the V380 Ori NE and HH 34 MMS cores,
because the ages of the protostars in these cores are not old enough
for the growth of millimeter-sized grains at the central region
and the subsequent transportation to a distance of $\sim$10,000 au.
The large grains should have been formed in the observed volume of the core.
The case of HH 147 MMS is less clear
because the protostar (N$^3$SK 50) is relatively evolved (Appendix).
The timescale issue is especially acute in the case of V380 Ori NE.
This core contains a single protostar (or a protobinary),
and its outflow timescale is only $\sim$6300 years (Section 4.2),
which is much shorter than the free-fall timescale.
Since the age of the protostar may be comparable to the outflow timescale,
the large grains should have been present already
at the onset of the protostellar collapse.
To grow these large grains,
the V380 Ori NE dense core must have spent a long time,
much longer than the free-fall timescale,
as a starless (i.e., pre-protostellar) core.
The protostellar age of HH 34 MMS (HOPS 188) is unclear
and may be in the order of 10$^5$ years (Appendix).
Therefore, the lifetime of some starless cores must be long enough
to form millimeter-sized dust grains.

The evolution of dense core is governed by the physical process
providing support against gravity.
Magnetically supported cores may form stars
in ambipolar diffusion timescales (Shu et al. 1987).
By contrast, gravoturbulent star formation theory suggests
that dense cores are transient objects supported by turbulence
and collapse rapidly to form stars in free-fall timescales
(Mac Low \& Klessen 2004; Klessen et al. 2005).
The dense cores containing large dust grains found in this study
seem to be counterexamples to the gravoturbulent theory.
Detailed study of these cores may provide information
on how they were supported during the pre-protostellar core stage.

Recent studies suggested
that dust grains can play important roles
in the structure and stability of dense cores
(Whitworth \& Bate 2002; Seo \& Youdin 2016; Bate \& Lor{\'e}n-Aguilar 2017).
If the starless stage of dense cores can last
long enough to produce large dust grains,
there are interesting implications on the evolution of the cores.
Such cores are likely to be in a quasistatic or marginally subcritical state,
because substantially supercritical or subcritical cores
may collapse or dissipate, respectively, in relatively short timescales.
Large grains can occupy a significant fraction of the total dust mass,
though they are small in numbers and invisible in extinction studies.
The drag force of gas becomes ineffective,
and the grains can be dynamically decoupled.
At a density of 10$^5$ cm$^{-3}$ and a temperature of 10 K,
grains larger than $\sim$160 $\mu$m are decoupled from gas.
Charged large grains are also decoupled from magnetic fields
as the gyration period becomes very long.
Because these grains are not supported
by gas pressure, turbulence, or magnetic fields,
they precipitate down and arrive at the center of the dense core
in about $\sqrt{2}$ free-fall timescale.
While completely decoupled grains oscillate or orbit around the center,
moderately decoupled ones would settle in the high-density region
(Bate \& Lor{\'e}n-Aguilar 2017).
As a result, the concentration of large grains
increases in the central region.
The high dust concentration also enhances cooling of gas.
Consequently, the density of the central region increases,
and the core becomes unstable,
which can trigger gravitational collapse.
In this picture, the lifetime of starless dense cores
can be affected or even limited by the production of large grains.
The change of stability condition related with the inward migration of dust
was studied by Whitworth \& Bate (2002),
in the context of external radiation pressure.
Hopkins (2014) discussed the collapse process of cores
with large dust-to-gas ratios.
Numerical simulations of large grains so far have investigated
dynamics or growth of grains in already unstable and collapsing cores
(Wong et al. 2016; Bate \& Lor{\'e}n-Aguilar 2017).
Simulations of large grains in quasistatic cores will be useful
in understanding the star formation activities in low-$\beta$ cores
such as V380 Ori NE, HH 34 MMS, and HH 147 MMS.
It is important to consider the precipitation,
concentration, and growth of grains together,
because these processes may affect each other.
Observational searches for starless cores with low $\beta$ values
will also be helpful in understanding the structure and evolution
of the cores containing large grains.

\subsection{Chemical Signatures of Formation History}

The old age of a dense core can leave some chemical signatures
owing to the depletion of molecules and grain-surface chemical reactions.
Though a quantitative analysis of the chemical age is not easy,
comparisons among dense cores can provide indications of relative age. 
Kang et al. (2015) found that the deuterium fractionation
of the V380 Ori NE core ([HDCO]/[H$_2$CO] = 0.12) is higher
than that of typical Class 0 protostellar cores ([HDCO]/[H$_2$CO] = 0.07),
which suggests that this core
has a history of high depletion of CO and H$_2$CO (Bacmann et al. 2003).

The V380 Ori NE outflow shows
the strongest CH$_3$OH thermal emission
among the 99 low-mass protostars in Orion surveyed by Kang et al. (2013).
The CH$_3$OH peak position (KLC 8)
corresponds to the brightest SiO outflow peak r6 (Figure 1).
This positional coincidence is consistent with the explanation
that the CH$_3$OH abundance is enhanced in protostellar outflows
through desorption or sputtering from grain surfaces
by hot gas in the shocked region
(Bachiller et al. 1995; Kristensen et al. 2010).
Interstellar CH$_3$OH molecules are mainly formed on grain surfaces
through hydrogenation of CO at low temperature (Fuchs et al. 2009).
Cuppen et al. (2009) found
that CH$_3$OH molecules form efficiently in cold dense cores
and dominate the ice layer of grain mantle with time.
The strong CH$_3$OH emission of V380 Ori NE
agrees with the low dust emissivity index,
in that this core may have spent a long time
in the pre-protostellar core stage.

\section{SUMMARY}

The star formation activities in the V380 Ori NE region
were investigated by observing
in the SiO $v$ = 0 $J$ = 1 $\rightarrow$ 0 line with VLA,
in the H$^{13}$CO$^+$ $J$ = 1 $\rightarrow$ 0,
C$^{34}$S $J$ = 2 $\rightarrow$ 1,
and CO $J$ = 1 $\rightarrow$ 0 lines with TRAO,
and in the $\lambda$ = 2.0, 1.4, 1.1, and 0.89 mm continuum with CSO.
The highly collimated SiO jet of V380 Ori NE
was imaged with a resolution of 1\farcs6,
and the detailed structure of the jet was examined.
The full extent of the V380 Ori NE bipolar outflow was measured
using the CO map covering a 10$'$ $\times$ 10$'$ region.
The continuum map covers a 60$'$ $\times$ 60$'$ region,
and the dust continuum spectra of several dense cores
were examined to investigate the dust properties.
The main results are summarized as follows.

1.
The SiO map shows a highly collimated bipolar jet,
and emission peaks along the jet were identified.
The SiO emission peaks show point-symmetric oscillation patterns
in both the position angle and the velocity,
when plotted as functions of distance from the driving source.
The point symmetry suggests that the cause of the outflow variability
is intrinsic to the driving source,
and the most likely explanation is the precession of the jet axis.
The length of the jet in the SiO image
is longer than a full cycle of the jet precession.

2.
The oscillation patterns of the precessing jet
were fitted with a set of sinusoidal functions
to derive the quantities describing the variability.
From the best-fit functions,
the wavelength of oscillation is 28$''$,
the oscillation amplitude of position angle is 3\fdg5,
the amplitude of velocity oscillation is 2.0 km s$^{-1}$,
and the mean line-of-sight velocity is 8.2 km s$^{-1}$
with respect to the systemic velocity.
These variability parameters were converted
to physical parameters describing the precessing jet,
assuming that the flow elements are on the surface of a circular cone
and have a uniform speed.
The derived flow speed is 35 km s$^{-1}$,
and the precession cone has an inclination angle of 13\fdg5
and a half-opening angle of 3\fdg4.

3.
The jet parameters provide useful information
on the evolutionary status of the V380 Ori NE system.
From the flow speed, inclination angle, and oscillation wavelength,
the jet precession period is 1600 years.
The full length of the outflow in the CO map
corresponds to $\sim$3.9 precession cycles,
and the dynamical timescale of the V380 Ori NE outflow is $\sim$6300 years.
Because the inclination angle was estimated explicitly,
the outflow timescale may be a good estimation of the protostellar age.
Considerations of the age and the flow speed suggest
that the mass of the protostar may be $\sim$0.02 $M_\odot$.
The outflow timescale and the mass estimate confirm
that the protostar is extremely young,
as previously inferred from the SED (Stutz et al. 2013).

4.
The jet precession suggests
that V380 Ori NE may be a protostellar binary system
with a misalignment between the disk and the orbital plane.
Two modes of tidal perturbation can be considered.
If the precession of disk is responsible
for the observed precession of the SiO jet,
the orbital period and separation of the binary system
may be $\sim$80 years and $\sim$6 au, respectively.
If the wobbling of disk is the responsible mechanism,
the corresponding values may be $\sim$3200 years and $\sim$70 au.
These estimates of binary separation suggest
that the presumed V380 Ori NE binary system
may have been formed through the disk instability process.

5.
The MUSIC continuum maps show several sources in the L1641 region.
Eight sources were identified using the 1.4 and 1.1 mm continuum maps.
Their spectral slopes suggest
that the detected flux densities come
almost entirely from the thermal dust emission.
Three dense cores, V380 Ori NE, HH 34 MMS, and HH 147 MMS,
show particularly shallow SED slopes.

6.
The SED of each dense core was fitted with a modified blackbody function,
assuming a power-law emissivity ($\kappa_\nu \propto \nu^\beta$).
In addition to the MUSIC data,
the 850 $\mu$m flux densities of Nutter \& Ward-Thompson (2007) were used.
The single-component dust emission models fit the data reasonably well.
The emissivity index $\beta$ and the mass of each core were derived
assuming a dust temperature of 20 K.
While the $\beta$ values of the majority of dense cores
are consistent with those of other dense cores in the literature,
V380 Ori NE, HH 34 MMS, and HH 147 MMS have unusually low values:
$\beta$ = 0.3, 0.7, and 0.7, respectively.
Considering the relatively large uncertainties,
these low $\beta$ values are tentative.
Because such low values ($\beta <$ 1) imply
that the maximum size of dust grains may be
at least a few times the wavelength observed (Draine 2006),
these three cores may contain a substantial amount
of millimeter-sized dust grains.

7.
The detected dense cores show a positive correlation
between the dust emissivity index and molecular line width.
The strength of turbulence may be related with the dust size distribution
by affecting the shattering and coagulation processes of dust grains.
The emissivity index has no strong correlation
with the bolometric luminosity or temperature of the protostars.

8.
Growing dust grains to a large size ($>$100 $\mu$m)
takes millions of years,
much longer than the free-fall timescale (Ormel et al. 2009).
By contrast, the age of the protostars
in V380 Ori NE and HH 34 MMS are shorter.
In particular, the protostellar age of V380 Ori NE,
inferred from the outflow timescale, is only about 6300 years,
which is much shorter than the free-fall timescale.
The stark difference between the grain growth timescale
and the protostellar age suggests
that these cores should already have produced large dust grains
during their starless core or pre-protostellar core stage of evolution.
Therefore, the lifetime of some starless dense cores
may be much longer than the free-fall timescale.
The large dust grains may be an interesting component of dense cores
because they are dynamically decoupled from gas
(Bate \& Lor{\'e}n-Aguilar 2017).
Because they can destabilize starless cores
by affecting the density structure and cooling process,
they may play an important role in the evolution of dense cores.

\acknowledgements

We thank the TRAO and the CSO staffs for their support.
NRAO is a facility of the National Science Foundation
operated under cooperative agreement by Associated Universities, Inc.
M. Kang was supported by Basic Science Research Program
through the National Research Foundation of Korea
funded by the Ministry of Science and ICT
(grant No. NRF-2015R1C1A1A01052160).
J.-E. Lee was supported by the Basic Science Research Program
through the National Research Foundation of Korea
(grant No. NRF-2015R1A2A2A01004769)
and the Korea Astronomy and Space Science Institute
under the R\&D program (Project No. 2015-1-320-18)
supervised by the Ministry of Science and ICT.

\appendix

\section{Individual Dense Cores}

\subsection{L1641N MMS}

The brightest source in the MUSIC maps is referred to as L1641N MMS
in this paper.
It is a complex of several sub-cores and protostars.
Stanke \& Williams (2007) imaged the core with a $\sim$1\farcs4 resolution
and identified sub-cores MM1--4.
Furlan et al. (2016) identified three protostars in this core.
The most deeply embedded one, HOPS 182,
is a Class 0 YSO with $L_{\rm bol}$ = 71.1 $L_\odot$
and located at the sub-core MM1.
The core shows relatively strong emission in the CH$_3$OH and CH$_3$CN lines
(Stanke \& Williams 2007; Kang et al. 2013).
Stanke \& Williams (2007) suggested that there is a hot corino around MM1.
There are at least four bipolar CO outflows
(Stanke \& Williams 2007; Nakamura et al. 2012)
and an active H$_2$O maser source (Kang et al. 2013).

L1641N MMS is the most extended source in the MUSIC maps.
It is elongated in the northeast--southwest direction,
with a position angle of $\sim$43$^\circ$
and a deconvolved size of $\sim$44$''$ (0.09 pc).
The emissivity index $\beta$ = 1.4 is near the median value
among the detected cores.
There is a secondary peak at $\sim$55$''$ northeast
of the main peak (Figure 10(c)).
Though there is no known protostar in the secondary peak,
the H$_2$O maser reported by Xiang \& Turner (1995)
is probably associated with it.

\subsection{Strom 11}

Strom 11 is a group of embedded near-infrared objects
associated with IRAS 05339--0626 (Strom et al. 1989; Chen et al. 1993).
There is a blueshifted CO outflow in this region
(Stanke \& Williams 2007; Nakamura et al. 2012).
Furlan et al. (2016) identified five protostars in this core,
all with relatively low (less than 2 $L_\odot$) luminosities.
Two of them (HOPS 173 and 380) are Class 0 YSOs.

The Strom 11 dense core has the steepest spectral slope
and the highest emissivity index ($\beta$ = 2.2) among the detected cores.
Even if a higher dust temperature is used for the calculation,
the emissivity index remains high.
Assuming $T_d$ = 40 K, for example,
the SED fit gives $\beta$ = 2.1.
This core does not seem to contain large dust grains.

\subsection{TUKH 74}

TUKH 74 is the only starless core among the detected cores.
The emissivity index $\beta$ = 1.4 is near the median value.
Considering that it does not have an internal heating source,
the dust temperature could be relatively low.
However, Ohashi et al. (2014) derived a kinetic temperature of 17 K
from the NH$_3$ data of Wilson et al. (1999).
Therefore, the assumed dust temperature of 20 K is reasonable.
In the grain model of Ricci et al. (2010),
$\beta$ = 1.4 implies a maximum grain size of $\sim$1 mm.
Considering the uncertainty,
it is unclear whether this starless core contains large dust grains.

\subsection{HH 34 MMS}

The core associated with the HH 34 jet
is referred to as HH 34 MMS in this paper.
There are two sub-cores that are not clearly separated by the MUSIC beams
(see Figure 14 of Johnstone \& Bally 2006).
The main core contains the driving source of HH 34,
and the secondary core contains IRS 5.
Furlan et al. (2016) identified three Class I protostars in this region.
HOPS 188 has $L_{\rm bol}$ = 18.8 $L_\odot$ and $T_{\rm bol}$ = 103 K
and corresponds to the driving source of the HH 34 jet.
Because the bolometric temperature is not much higher
than the Class 0/I boundary (70 K),
the age of this protostar may be somewhat older
than the typical lifetime of the Class 0 phase
($\sim$10$^5$ years; Evans et al. 2009).
The prominent HH 34 jet has been studied extensively
(Reipurth et al. 1986, 2002; Rodr{\'\i}guez \& Reipurth 1996;
Nisini et al. 2016).
The jet extends to a parsec-scale outflow
that has a dynamic age of $\sim$10,000 years
(Bally \& Devine 1994; Devine et al. 1997).
Considering both the YSO classification and outflow timescale,
the protostellar age of HOPS 188 may be on the order of 10$^5$ years.
HOPS 189 corresponds to IRS 5
and has $L_{\rm bol}$ = 1.3 $L_\odot$.
HOPS 190 corresponds to Object 22 of Reipurth et al. (1986, 2002)
and has $L_{\rm bol}$ = 0.4 $L_\odot$.

Most of the emission detected by MUSIC comes from the main core.
The low emissivity index $\beta$ = 0.7 suggests
that this core contains millimeter-sized dust grains.

\subsection{V380 Ori NE}

V380 Ori NE is a relatively isolated core.
Furlan et al. (2016) identified a protostar in this core:
HOPS 169 is a Class 0 YSO
with $L_{\rm bol}$ = 3.9 $L_\odot$ and $T_{\rm bol}$ = 33 K.
The star formation activities of V380 Ori NE are described in the main text.

V380 Ori NE has the lowest dust emissivity index, $\beta$ = 0.3,
among the detected cores,
which suggests that it contains millimeter-sized dust grains.
Davis et al. (2000) suggested
that the submillimeter emission from the extended  part of the core
probably comes from dust heated by the outflow.
If the dust temperature is higher, however,
it would make $\beta$ even lower.
Assuming $T_d$ = 30 K, for example, the SED fit gives $\beta$ = 0.2.

\subsection{HH 147 MMS}

HH 147 MMS is the dense core containing N$^3$SK 50,
the driving source of the HH 147 outflow (Nakajima et al. 1986;
Eisl{\"o}ffel et al. 1994; Chini et al. 1997, 2001; Choi \& Zhou 1997).
The nature of this YSO is unclear.
It was known to be a rare type of T Tauri star
exhibiting a P Cygni profile in H$\alpha$
(Strom et al. 1989; Corcoran \& Ray 1995).
By contrast, Chini et al. (2001) noted
that the bolometric-to-submillimeter luminosity ratio of the core
indicates a Class 0 YSO.
Furlan et al. (2016) classified its counterpart, HOPS 166,
as a flat-spectrum YSO with $L_{\rm bol}$ = 15.5 $L_\odot$
and $T_{\rm bol}$ = 457 K.
Moro-Mart{\'\i}n et al. (1999) detected a CO outflow.

The emissivity index is low ($\beta$ = 0.7),
but the uncertainty is relatively large.
This core probably contains millimeter-sized dust grains.

\subsection{HH 1--2 MMS 2}

The HH 1--2 MMS 2 core in the MUSIC maps includes
MMS 2 and MMS 3 of Chini et al. (1997, 2001).
There are two compact sources, VLA 3 and VLA 6,
detected in centimeter continuum (Rodr{\'\i}guez et al. 1990, 2000).
VLA 3 corresponds to HOPS 168,
a Class 0 YSO with $L_{\rm bol}$ = 48.1 $L_\odot$ (Furlan et al. 2016).
This protostar displays H$_2$O maser activities
(Lo et al. 1975; Kang et al. 2013).
There is a powerful CO outflow in this region
(Chernin \& Masson 1995; Choi \& Zhou 1997; Moro-Mart{\'\i}n et al. 1999).
VLA 6 is a time-variable source
and probably a magnetically active YSO (Rodr{\'\i}guez et al. 2000).
Furlan et al. (2016) identified yet another protostar
located between MMS 1 and MMS 2:
HOPS 167 corresponds to source 21 of Kwon et al. (2010)
and has $L_{\rm bol}$ = 0.2 $L_\odot$.

HH 1--2 MMS 2 has a relatively high dust emissivity index ($\beta$ = 1.9).
This core does not seem to contain large dust grains.

\subsection{HH 1--2 MMS 1}

HH 1--2 MMS 1 is the dense core containing the driving source
of the archetypal Herbig-Haro objects HH 1 and HH 2
(Chini et al. 1997, 2001).
The optical jet, HH objects, and their environment
have been studied extensively (Strom et al. 1989; Reipurth et al. 1993;
Eisl{\"o}ffel et al. 1994; Corcoran \& Ray 1995; Kwon et al. 2010).
There are three compact sources, VLA 1/2/4,
detected in centimeter continuum (Rodr{\'\i}guez et al. 2000).
VLA 1 corresponds to HOPS 203,
a Class 0 YSO with $L_{\rm bol}$ = 20.4 $L_\odot$ (Furlan et al. 2016).
It is the driving source of the HH 1--2 outflow.
The CO outflow associated with the HH 1--2 jet is relatively weak
(Chernin \& Masson 1995; Choi \& Zhou 1997; Moro-Mart{\'\i}n et al. 1999).
The H$_2$O maser activities reported by Kang et al. (2013)
is probably associated with a VLA 1 or HH 1 jet.
VLA 2 probably drives the HH 144 outflow, but its nature is unclear
(Reipurth et al. 1993; Eisl{\"o}ffel et al. 1994).
VLA 4 corresponds to HOPS 165,
a Class I YSO with $L_{\rm bol}$ = 3.4 $L_\odot$ (Furlan et al. 2016).
It is probably the driving source of the HH 146 outflow
(Reipurth et al. 1993).

The HH 1--2 MMS 1 dense core
has a relatively high dust emissivity index ($\beta$ = 1.6).
The MMS 1 core does not seem to contain large dust grains.

The three cores (HH 1--2 MMS 1 and 2, and HH 147 MMS)
are spatially close to each other,
and their star-forming activities may be considered
as a single cluster-forming event.
Kwon et al. (2010) found that the magnetic fields in this region
are regular and straight,
but the outflow orientations are essentially random.
This contrast between magnetic fields and outflows suggests
that the star formation in this region is mainly driven by turbulence.
As predicted by the gravoturbulent star formation theory,
the ages of the HH 1--2 MMS 1 and 2 cores are probably too short
to form large dust grains.

\end{document}